\documentclass[twocolumn,tighten]{aastex63}

\usepackage{natbib}
\usepackage{amsmath}
\usepackage{graphicx}
\usepackage[hyphens]{url}
\hypersetup{breaklinks=true}

\begin{document}

\title{Hot Dust-Obscured Galaxies with Excess Blue Light}

\author{R.J.~Assef}
\affiliation{N\'ucleo de Astronom\'ia de la Facultad de Ingenier\'ia y
  Ciencias, Universidad Diego Portales, Av. Ej\'ercito Libertador
  441, Santiago, Chile.}

\author{M.~Brightman}
\affiliation{Cahill Center for Astrophysics, California Institute of
  Technology, 1216 East California Boulevard, Pasadena, CA 91125, USA}

\author{D.J.~Walton}
\affiliation{Institute of Astronomy, University of Cambridge,
  Madingley Road, Cambridge CB3 0HA, UK}

\author{D.~Stern}
\affiliation{Jet Propulsion Laboratory, California Institute of
  Technology, 4800 Oak Grove Drive, Pasadena, CA 91109, USA}

\author{F.E.~Bauer}
\affiliation{Instituto de Astrof\'isica, Facultad de F\'isica,
  Pontificia Universidad Cat\'olica de Chile, 306, Santiago 22, Chile}
\affiliation{Millennium Institute of Astrophysics (MAS), Nuncio
  Monse\~nor S\'otero Sanz 100, Providencia, Santiago, Chile}
\affiliation{Space Science Institute, 4750 Walnut Street, Suite 205,
  Boulder, Colorado 80301}

\author{A.W.~Blain}
\affiliation{Physics \& Astronomy, University of
  Leicester, 1 University Road, Leicester LE1 7RH, UK}

\author{T.~D\'iaz-Santos}
\affiliation{N\'ucleo de Astronom\'ia de la Facultad de Ingenier\'ia y
  Ciencias, Universidad Diego Portales, Av. Ej\'ercito Libertador
  441, Santiago, Chile.}
\affiliation{Chinese Academy of Sciences South America Center for Astronomy (CASSACA), National Astronomical Observatories, CAS, Beijing 100101, China}
\affiliation{Institute of Astrophysics, Foundation for Research and Technology-Hellas (FORTH), Heraklion, GR-70013, Greece}

\author{P.R.M.~Eisenhardt}
\affiliation{Jet Propulsion Laboratory, California Institute of
  Technology, 4800 Oak Grove Drive, Pasadena, CA 91109, USA}

\author{R.C.~Hickox}
\affiliation{Department of Physics and Astronomy, Dartmouth College,
  6127 Wilder Laboratory, Hanover, NH 03755, USA}

\author{H.D.~Jun}
\affiliation{School of Physics, Korea Institute for Advanced Study, 85
  Hoegiro, Dongdaemun-gu, Seoul 02455, Korea}

\author{A.~Psychogyios}
\affiliation{Department of Physics, University of Crete, GR-71003 Heraklion, Greece}
\affiliation{Institute of Astrophysics, FORTH, Heraklion, GR-70013, Greece}

\author{C.-W.~Tsai}
\affiliation{National Astronomical Observatories, Chinese Academy of
  Sciences, 20A Datun Road, Chaoyang District, Beijing, 100012,
  People's Republic of China}

\author{J.W.~Wu}
\affiliation{National Astronomical Observatories, Chinese Academy of
  Sciences, 20A Datun Road, Chaoyang District, Beijing, 100012,
  People's Republic of China}

\begin{abstract}

  Hot Dust-Obscured Galaxies (Hot DOGs) are among the most luminous
  galaxies in the Universe. Powered by highly obscured, possibly
  Compton-thick, active galactic nuclei (AGNs), Hot DOGs are
  characterized by SEDs that are very red in the mid-IR yet dominated
  by the host galaxy stellar emission in the UV and optical. An
  earlier study identified a sub-sample of Hot DOGs with significantly
  enhanced UV emission. One target, W0204--0506, was studied in detail
  and, based on {\it{Chandra}} observations, it was concluded that the
  enhanced emission was most likely due to either extreme unobscured
  star-formation (${\rm SFR}>1000~M_{\odot}~\rm yr^{-1}$) or to light
  from the highly obscured AGN scattered by gas or dust into our line
  of sight. Here, we present a follow-up study of W0204--0506 as well
  as two more Hot DOGs with excess UV emission. For the two new
  objects we obtained {\it{Chandra}}/ACIS-S observations, and for all
  three targets we obtained {\it{HST}}/WFC3 F555W and F160W
  imaging. The analysis of these observations, combined with
  multi-wavelength photometry and UV/optical spectroscopy suggest that
  UV emission is most likely dominated by light from the central
  highly obscured, hyper-luminous AGN that has been scattered into our
  line of sight, by either gas or dust. We cannot decisively rule out,
  however, that star-formation or a second AGN in the system may
  significantly contribute to the UV excess of these targets.

\end{abstract}


\section{Introduction}

Hot Dust-Obscured Galaxies \citep[Hot DOGs;][]{eisenhardt12,wu12} are
some of the most luminous galaxies in the Universe, with bolometric
luminosities $L_{\rm bol}>10^{13}~L_{\odot}$ and a significant
fraction with $L_{\rm bol}>10^{14}~L_{\odot}$
\citep{wu12,tsai15}. Discovered by the Wide-field Infrared Survey
Explorer \citep[WISE;][]{wright10}, Hot DOGs are characterized by very
red mid-IR colors and spectral energy distributions (SEDs) that peak
at rest-frame $\sim 20\mu\rm m$. This implies that Hot DOGs are
powered by highly obscured, hyper-luminous AGN that dominate the SED
from the mid- to the far-IR
\citep{eisenhardt12,wu12,wu14,fan16a,diaz16,tsai18}. As expected from
their luminosities, Hot DOGs are rare, with one object every 31$\pm$4
deg$^2$. Yet their number density is comparable to that of similarly
luminous unobscured quasars \citep{assef15} and of heavily reddened
type 1 quasars \citep{banerji15}.

X-ray studies have shown that the obscuration of the central engine in
Hot DOGs is very high, with column densities ranging from somewhat
below to above the Compton-thick limit \citep[i.e., $N_{\rm
    H}>1.5\times 10^{24}~\rm
  cm^{-2}$][]{stern14,piconcelli15,assef16,ricci17,vito18}. As the AGN
emission is highly obscured, the host galaxy is observable at
rest-frame UV, optical and near-IR wavelengths. A study of their SEDs
by \citet{assef15} showed that their stellar masses, as derived from
their rest-frame near-IR luminosities, imply that either the
super-massive black holes (SMBHs) are accreting well above the
Eddington limit, or that their SMBH masses ($M_{\rm BH}$) are well
above the local relations between $M_{\rm BH}$ and the mass of the
spheroidal component of the host galaxy \citep[see,
  e.g.,][]{magorrian98,bennert11}. Indeed, recent results by
\citet{wu18} and \citet{tsai18} suggest that Hot DOGs are radiating at
or above the Eddington limit, which in turn suggests that Hot DOGs are
likely experiencing strong AGN feedback that could easily affect the
whole host galaxy and its immediate environment. Indeed,
\citet{diaz16} presented a study of the [C\,{\sc ii}] 157.7$\mu$m
emission line in the highest luminosity Hot DOG, and possibly the most
luminous galaxy known, WISEA J224607.56--052634.9
\citep[W2246--0526;][]{tsai15}, and determined based on the
emission-line kinematics that the central gas of the host galaxy is
likely undergoing an isotropic outflow event. Further ionized gas
outflow signatures have been observed in the optical narrow emission
lines of some other Hot DOGs \citep{wu18,jun18}, supporting the
presence of strong AGN feedback in the ISM of these targets.

\citet[][also see \citealt{eisenhardt12,tsai15,tsai18}]{assef15}
showed that the UV through mid-IR SED of the majority of Hot DOGs
(specifically ``W12--drops'' with $z>1$) can be well modeled as a
combination of a star-forming galaxy that dominates the optical/UV
emission, and a luminous, obscured AGN that dominates the mid-IR SED
and the bolometric luminosity of the system. However, this is not the
case for all Hot DOGs. In a later work, \citet[][A16
  hereafter]{assef16} presented a small sample of eight Hot DOGs whose
optical/UV emission is not well modeled by a star-forming galaxy, but
instead needs a second, unobscured AGN component that is only
$\sim$1\% as luminous as the obscured component. A16 posited that the
SED could be explained by three different scenarios: i) that the
UV/optical emission is dominated by leaked or scattered light from the
hyper-luminous, highly obscured AGN; ii) that the system is a dual
quasar, with a more luminous, highly obscured quasar and a less
luminous, unobscured one; and iii) that the system is undergoing an
extreme star-formation event with little dust obscuration such that
the broad-band UV/optical SED is similar to that of an AGN.

One of these objects, WISEA J020446.13--050640.8 (W0204--0506
hereafter), was serendipitously observed by the {\it{Chandra X-ray
    Observatory}} as part of the Large-Area Lyman Alpha survey
\citep[LALA;][]{rhodes00}. A16 studied this object in detail using
these observations along with broad-band SED and optical spectroscopic
observations. A16 determined that the X-ray spectrum of W0204--0506 is
consistent with a single, hyper-luminous, highly absorbed AGN ($\log
L_{2-10~\rm keV}/\rm erg~\rm s^{-1}=44.9^{+0.86}_{-0.14}, $ $N_{\rm H}
= 6.3^{+8.1}_{-2.1}\times 10^{23}~\rm cm^{-2}$), and highly
inconsistent with a secondary, unobscured AGN with the luminosity
necessary to explain the optical/UV emission. Instead, A16 found that
the UV/optical continuum was better explained by a starburst with a
star-formation rate $\gtrsim 1000~M_{\odot}~\rm yr^{-1}$, or by
scattered light from the hyper-luminous, highly obscured central
engine. While star-formation rates (SFRs) $\gtrsim 1000~M_{\odot}~\rm
yr^{-1}$ are routinely found through far-IR/sub-mm observations of
highly obscured systems such as SMGs and ULIRGs, rates above $\sim
300~M_{\odot}~\rm yr^{-1}$ have never been observed through UV/optical
wavelengths in Lyman break galaxies, which have the strongest
UV/optical star-formations measured \citep{barger14}. Due to the large
SFR needed to explain the optical/UV SED of this object as a
starburst, A16 favored the scattered AGN-light scenario.

In this paper we present {\it{Hubble Space Telescope}} ({\it{HST}})
observations of W0204--0506 to further explore its nature, and we
explore in detail two more Blue-Excess Hot DOGs (BHDs), WISE
J022052.12+013711.6 (W0220+0137 hereafter) and WISE
J011601.41-050504.0 (W0116--0505 hereafter), using {\it{HST}} and
{\it{Chandra}} observations as well as optical spectroscopy and
broad-band UV through mid-IR SEDs. In \S\ref{sec:observations} we
present the sample studied here as well as the different observations
available for each target, and describe the SED modeling done by A16
to identify these unusual objects. In \S\ref{sec:xray_model} we
discuss the modeling of the {\it{Chandra}} X-ray observations. In
\S\ref{sec:discussion} we present a detailed discussion of the source
of the excess blue emission, analyzing each possible scenario in light
of the available observations. Our conclusions are summarized in
\S\ref{sec:summary}. Throughout the article all magnitudes are
presented in their natural system unless otherwise stated, namely AB
for $ugriz$ and Vega for all the rest. We assume a concordance flat
$\Lambda$CDM cosmology with $H_0=70~\rm km~\rm s^{-1}~\rm Mpc^{-1}$,
$\Omega_{\Lambda}=0.7$ and $\Omega_{\rm M} = 0.3$. For all quantities
derived from X-ray spectra, we quote 90\% confidence interval, while
for all other quantities we quote 68.3\% confidence intervals instead.

\section{Sample and Observations}\label{sec:observations}

\subsection{Blue-Excess Hot DOGs}\label{ssec:bhds}

A16 identified 8 BHDs from a sample of 36 Hot DOGs with W4$<$7.2~mag,
spectroscopic redshifts $z>1$ and {\it{ugriz}}
{\tt{modelMag}}\footnote{\url{http://www.sdss.org/dr12/algorithms/magnitudes/\#mag\_model}}
photometry in the SDSS DR12 database with $S/N>3$ in at least one of
the SDSS bands. This spectroscopic sample is biased towards optical
emission, and after considering the selection effects, A16 estimated
BHDs could comprise as much as 8\% of the Hot DOG population with
W4$<$7.2~mag, although most likely a smaller fraction when considering
fainter W4 fluxes.

To select this sample, A16 started by modeling the SEDs of the
aforementioned 36 Hot DOGs using the galaxy and AGN SED templates and
modeling algorithm of \citet{assef10}, following the prescription
presented by \citet{assef15}. In short, the broad-band SED of any
given object was modeled as a linear, non-negative combination of four
empirically derived SED templates: an ``E'' template, which resembles
the SED of an old stellar population, an ``Sbc'' template, which
resembles the SED of an intermediately star-forming galaxy, an ``Im''
template, which resembles a local starburst galaxy, and a type 1 AGN
template. The reddening of the AGN template, parametrized by $E(B-V)$,
was also fit for, assuming $R_V=3.1$ and a reddening law that follows
that of the SMC at short wavelengths but that of the Milky Way at
longer wavelengths. A single IGM absorption strength was also fit for
all templates when needed \citep[see][A16 for
  details]{assef10,assef15}. Using this approach A16 modeled the SED
of each object in the following broad bands: the {\it{ugriz}} SDSS
DR12 {\tt{modelMag}} photometry, {\it{Spitzer}}/IRAC [3.6] and [4.5]
photometry from \citet{griffith12}, and the WISE W3 and W4 photometry
from the WISE All-Sky Data Release \citep{cutri12}. Additionally,
whenever possible, A16 used the J, Ks and deeper $r$-band imaging
presented by \citet{assef15}. For the three sources considered in this
article, the deeper $r$-band imaging was obtained using the 4.1m
Southern Astrophysical Research Telescope (SOAR) with the SOAR Optical
Imager (SOI). For W0116--0505, images were obtained with an exposure
time of 3$\times$600~s on the night of UT 2013 August 28. For the
other two sources, the images were obtained on UT 2011 November 20,
with exposure times of 3$\times$500~s for W0204--0506 and of
2$\times$500~s for W0220+0137. In all cases the images were reduced
following standard procedures, and the photometric calibration was
performed by comparing bright stars in each field with their
respective SDSS magnitudes. The details of the NIR imaging can be
found in \citet{assef15}. All magnitudes are shown in Table
\ref{tab:phot}.

\begin{deluxetable*}{l c c c}

  \tablecaption{Photometric Data\label{tab:phot}}

  \tablehead{
    \colhead{WISE ID} &
    \colhead{J011601.41--050504.0} &
    \colhead{J020446.13--050640.8} &
    \colhead{J022052.12+013711.6}
  }

  \tabletypesize{\small}
  \tablewidth{0pt}
  \tablecolumns{4}

  \startdata
  SDSS $u$             & 23.571$\pm$0.685     & 23.004$\pm$0.600     & 23.470$\pm$0.587    \\
  SDSS $g$             & 21.464$\pm$0.054     & 22.660$\pm$0.166     & 21.779$\pm$0.059    \\
  F555W                & 21.679$\pm$0.019     & 22.441$\pm$0.047     & 21.772$\pm$0.018    \\
  SDSS $r$             & 21.383$\pm$0.054     & 22.488$\pm$0.234     & 21.841$\pm$0.086    \\
  SOI $r$              & 21.515$\pm$0.078     & 22.357$\pm$0.166     & 21.803$\pm$0.047    \\
  SDSS $i$             & 21.740$\pm$0.094     & 21.797$\pm$0.175     & 22.060$\pm$0.132    \\
  SDSS $z$             & 21.368$\pm$0.257     & 22.026$\pm$0.667     & 21.607$\pm$0.273    \\
  $J$                  & \nodata              & 20.768$\pm$0.216     & 20.790$\pm$0.149    \\
  F160W                & 20.648$\pm$0.007     & 20.390$\pm$0.007     & 21.077$\pm$0.010    \\
  $Ks$                 & \nodata              & \nodata              & 18.604$\pm$0.117    \\
  W1                   & 17.130$\pm$0.184     & 17.343$\pm$0.115     & 17.875$\pm$0.225    \\
  $[$3.6$]$            & 16.800$\pm$0.040     & 17.182$\pm$0.056     & 17.722$\pm$0.091    \\
  $[$4.5$]$            & 15.725$\pm$0.021     & 16.340$\pm$0.033     & 16.806$\pm$0.051    \\
  W2                   & 15.564$\pm$0.156     & 16.103$\pm$0.158     & 16.575$\pm$0.253    \\
  W3                   & 10.213$\pm$0.059     & 10.245$\pm$0.056     & 10.512$\pm$0.075    \\
  W4                   & \phn7.014$\pm$0.084  & \phn7.062$\pm$0.090  & \phn7.076$\pm$0.092 \\
  \enddata

\end{deluxetable*}

A16 found that the approach described above was not able to accurately
model the UV/optical emission for a fraction of their sample, which
were significantly bluer than allowed by the SED templates. They
identified eight objects for which an additional, secondary AGN
component with independent normalization and reddening provided a
significant improvement in $\chi^2$ to the best-fit SED model. A16
presented a detailed study of the properties of one of these targets:
W0204--0506. Here we study an additional one of these eight targets,
W0220+0137, as well as another very similar target, W0116--0505. The
W1$=$17.13$\pm$0.18~mag of W0116--0505 is slightly brighter than the
formal Hot DOG selection limit \citep[W1$>$17.4;][]{eisenhardt12} and
hence it was excluded from the final list presented by A16 despite
meeting all other selection criteria. We find there is only a 2.7\%
probability that the improvement in $\chi^2$ by the secondary AGN
component is spurious for this source. A16 argued that these
probabilities are likely overestimated and hence conservative, as the
F-test used to estimate them does not take into account the
constraints provided by the non-negative requirement of the linear
combination of templates for the best-fit model.

\begin{figure}
  \begin{center}
    \plotone{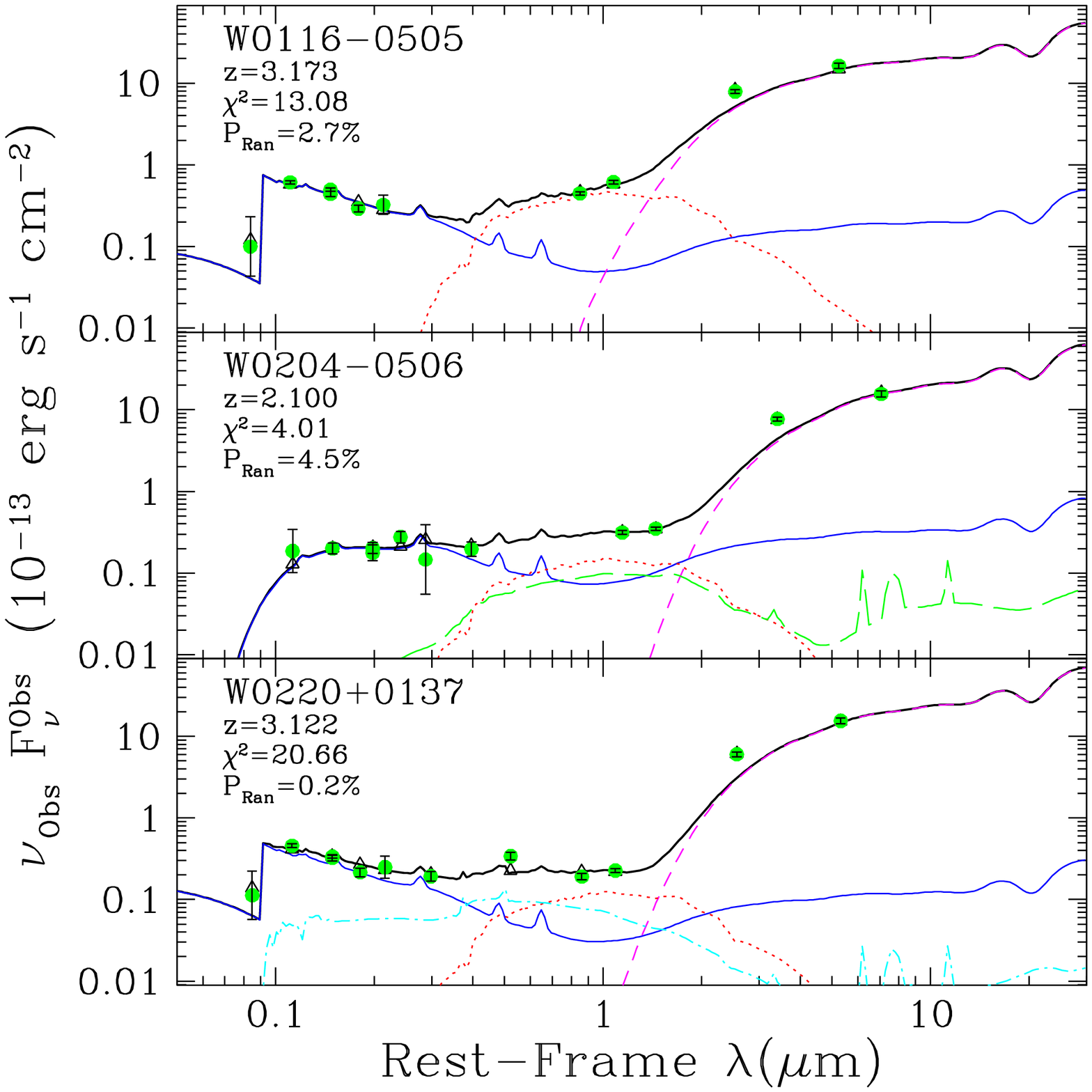}
    \caption{UV through mid-IR SEDs of the three BHDs discussed in
      this study, identified by their SEDs as discussed in the
      text. We only show the photometric data used to identify these
      targets as BHDs by A16. The green solid points show the observed
      flux densities in the photometric bands discussed in
      \S\ref{ssec:bhds}. The solid black line shows the best-fit SED
      model to the photometric data points, that consists of a
      non-negative linear combination of a primary luminous, obscured
      AGN (dashed magenta line), a secondary less luminous, unobscured
      or mildly obscured AGN (solid blue line), an old stellar
      population (dotted red line), an intermediate stellar population
      (dashed green line), and a young stellar population (cyan
      dotted-dashed line). The open triangles show the predicted flux
      density for each photometric band based on the best-fit SED
      model. For each object we indicate also the redshift and the
      probability $P_{\rm Ran}$ that the improvement in $\chi^2$
      gained from adding the secondary AGN component is spurious.}
    \label{fg:seds}
  \end{center}
\end{figure}

The broad band SEDs as well as best-fit SED models of the three targets are
shown in Figure \ref{fg:seds}. We note that the SED of W0204--0506 differs
slightly from that presented by A16 as, for consistency with the other two
sources, the SED presented here only uses the SDSS DR12 bands in the UV/optical
instead of the deeper imaging of \citet{finkelstein07}. Table \ref{tab:pars}
shows, for each target, the best-fit $E(B-V)$ to both the primary and secondary
AGN components and Figure \ref{fg:seds} shows the best-fit SED models.
The table also shows the reddening-corrected monochromatic luminosities at
6$\mu$m, $L_{6\mu\rm m}$, calculated from the template fit to each AGN
component. The uncertainties for the parameters shown in Table
\ref{tab:pars} have been estimated using a Monte-Carlo method following a
similar prescription to that used by A16. For each object we first apply a
scaling factor to the photometric uncertainties such that the best-fit SED model
has a reduced $\chi^2$ ($\chi^2_{\nu}$) of 1. We then create 1,000 realizations
of the observed SED of each object by re-sampling its photometry according to
these scaled uncertainties and assuming a Gaussian distribution. We
fit each of the 1,000 simulated SEDs and compile the distribution of each
parameter.  We assign the uncertainties to the 68.3\% intervals of these
distributions around the values of the best-fit model.

\begin{deluxetable*}{l c c c c c c c c c}

  \tablecaption{Best-fit SED Parameters\label{tab:pars}}

  \tablehead{
    \colhead{} &
    \colhead{} &
    \colhead{} &
    \multicolumn{2}{c}{Primary AGN}&
    \colhead{} &
    \multicolumn{2}{c}{Secondary AGN}&
    \colhead{} &
    \colhead{}\\
    \colhead{Object}&
    \colhead{Redshift}&
    \colhead{} &
    \colhead{$\log L_{6\mu\rm m}$} &
    \colhead{$E(B-V)$} &
    \colhead{} &
    \colhead{$\log L_{6\mu\rm m}$} &
    \colhead{$E(B-V)$} &
    \colhead{} &
    \colhead{$P_{\rm ran}$}\\
    \colhead{} &
    \colhead{} &
    \colhead{} &
    \colhead{($\rm erg~\rm s^{-1}$)} &
    \colhead{(mag)} &
    \colhead{} &
    \colhead{($\rm erg~\rm s^{-1}$)} &
    \colhead{(mag)} &
    \colhead{} &
    \colhead{($10^{-2}$)}
  }

  \tabletypesize{\small}
  \tablewidth{0pt}
  \tablecolumns{12}

  \startdata
  W0116$-$0505 & 3.173 & & 47.24$^{+0.17}_{-0.11}$ & \phn4.24$^{+2.71}_{-1.23}$ & & 45.18$^{+0.04}_{-0.03}$ & 0.00$^{+0.02}_{-0.00}$ & & 2.7\\
  W0204$-$0506 & 2.100 & & 46.87$^{+0.03}_{-0.08}$ &    10.00$^{+1.74}_{-2.06}$ & & 44.98$^{+0.04}_{-0.22}$ & 0.10$^{+0.00}_{-0.05}$ & & 4.5\\
  W0220$+$0137 & 3.122 & & 47.33$^{+0.16}_{-0.16}$ & \phn7.33$^{+2.67}_{-2.32}$ & & 44.96$^{+0.08}_{-0.11}$ & 0.00$^{+0.02}_{-0.00}$ & & 0.2\\
  \enddata

  \tablecomments{The uncertainties have been derived through a Monte Carlo process, as described in the text, and hence do not capture the possible systematic uncertainties described at the end of \S\ref{ssec:bhds}.}

\end{deluxetable*}

As the set of templates used is limited, we warn the reader that we may have
neglected significant systematic uncertainties. For example, Figure
\ref{fg:seds} shows that the best-fit SED models have significant contribution
from the old stellar population template, which could be at odds with the young
age of the Universe at the redshift of these objects. However, it should be
noted that since in the \citet{assef10} algorithm SEDs are modeled by linear
combinations of three empirical templates, stellar population ages should not be
associated to any specific template, but rather to the combination of them.
Furthermore, the host emission is only being constrained by the rest-frame NIR
and hence is insensitive to star-formation history of the underlying stellar
population. This will naturally lead to a significant degeneracy in the host
galaxy SED modeling.  As a test, we re-fit the SEDs without allowing old stellar
populations. The best-fit parameters listed in Table 2 are all within the
estimated uncertainties, and the $P_{\rm Ran}$ values all become somewhat
smaller (by factors of $\sim$3), indicating that our results are qualitatively
unaffected. Similarly, systematics in determining the AGN luminosities may also
have been neglected. A detailed analysis of the dust emission SED of Hot DOGs,
including data from {\it{Herschel}}, will be presented elsewhere and will
elucidate this issue further \citep{tsai20}.

\subsection{Optical Spectra}

\begin{figure}
  \begin{center}
    \plotone{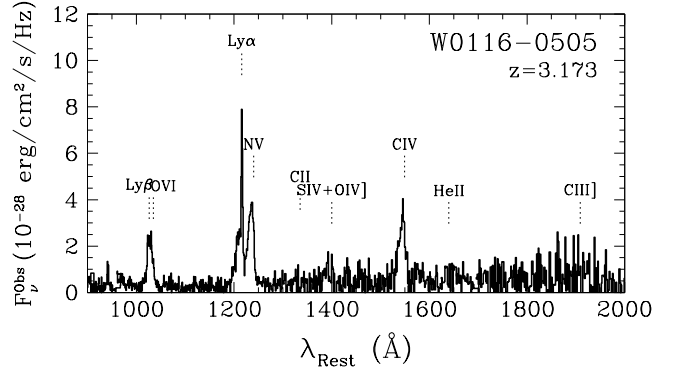}
    \caption{Optical spectrum of W0116--0505 from SDSS.}
    \label{fg:W0116_spec}
  \end{center}
\end{figure}

\begin{figure}
  \begin{center}
    \plotone{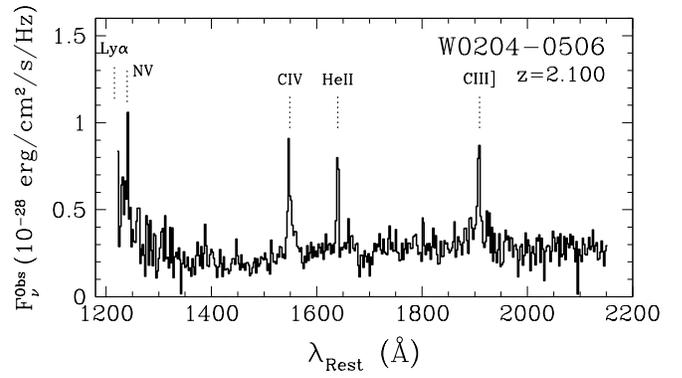}
    \caption{Optical spectrum of W0204--0506, obtained with the GMOS-S
      instrument at the Gemini South Observatory.}
    \label{fg:W0204_spec}
  \end{center}
\end{figure}

\begin{figure}
  \begin{center}
    \plotone{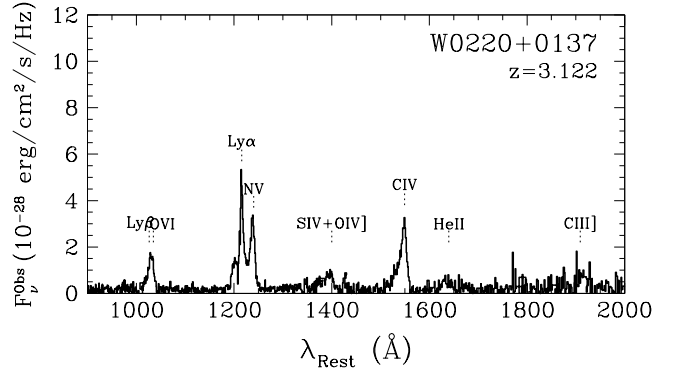}
    \caption{Optical spectrum of W0220+0137 from SDSS.}
    \label{fg:W0220_spec}
  \end{center}
\end{figure}

The optical spectra of W0116--0505, W0204--0506 and W0220+0137 are
presented in Figures \ref{fg:W0116_spec}, \ref{fg:W0204_spec} and
\ref{fg:W0220_spec}, respectively. For W0116--0505 and W0220+0137, the
spectra were obtained from SDSS. The optical spectrum of W0204--0506
was obtained using the GMOS-S spectrograph on the Gemini South
telescope on UT 2011 November 27 using a longslit with a width of
1.5\arcsec\ as well. These observations have been previously presented
by A16, and we refer the reader to that study for further details on
these observations. The spectra of W0116$-$0505 and W0220+0137 show
broad high ionization emission lines typically observed in luminous
quasars. W0204--0506, on the other hand, has a spectrum more similar
to obscured, or type~2, quasars such as high-redshift radio galaxies
\citep[e.g.,][]{stern99,debreuck01} and other radio-quiet (or
non-radio-selected) quasars at high redshift
\citep[e.g.,][]{stern02,hainline12,alexandroff13}. Single Gaussian
fits to the C\,{\sc iv} emission line, following the prescription of
\citet[][and references therein]{assef11} to fit the continuum and
define the spectral region on which to fit the emission line, have
FWHM of approximately 2800~$\rm km~\rm s^{-1}$ and 3500~$\rm km~\rm
s^{-1}$ respectively for W0116--0505 and W0220+0137. Based on these
emission lines we measure a redshift of $z=3.173\pm 0.002$ for
W0116--0505, and $z=3.122\pm 0.002$ for W0220+0137. In particular both
spectra show blended Ly$\beta$ and O\,{\sc vi} emission
features. W0204--0506 is at a significantly lower redshift of
$z=2.100\pm 0.002$, and hence we cannot determine if these emission
lines are present in the spectrum, as they fall shortwards of the
atmospheric UV cut-off.

\smallskip
\subsection{HST Observations}\label{ssec:hst_data}

A joint program between {\it{Chandra}} and {\it{HST}} was approved
during {\it{Chandra}} Cycle 17 (PID 17700696) to obtain {\it{HST}}
imaging in two bands of all three targets and obtain
{\it{Chandra}}/ACIS-S observations of W0116--0505 and
W0220+0137. These targets were selected for having some of the
clearest blue excess emission in terms of the $\chi^2$ improvement,
and for having some of the highest expected count rates in ACIS-S. The
archival {\it{Chandra}}/ACIS-I observations for W0204--0506 presented
by A16 are sufficient to accomplish our science goals, so no further
observations were requested. In this section we focus on the
{\it{HST}} observations, while the {\it{Chandra}} observations are
described in the next section.

\begin{figure*}
  \begin{center}
    \includegraphics*[width=0.30\textwidth]{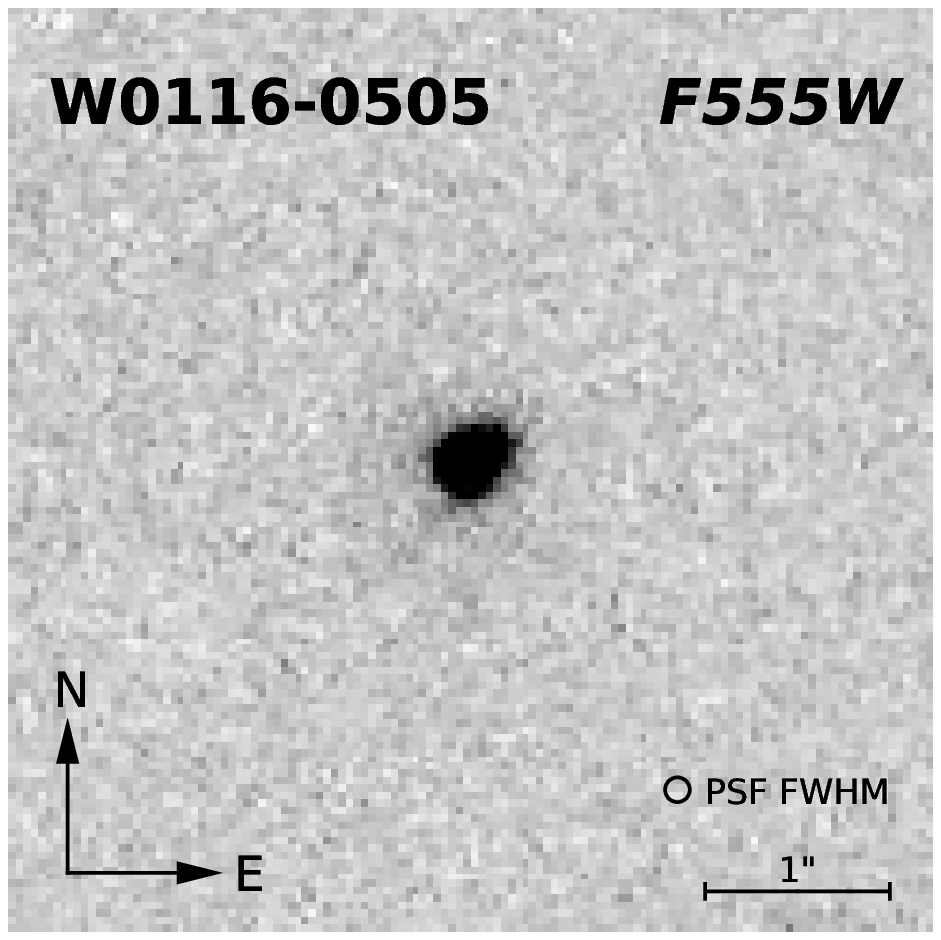}
    \includegraphics*[width=0.30\textwidth]{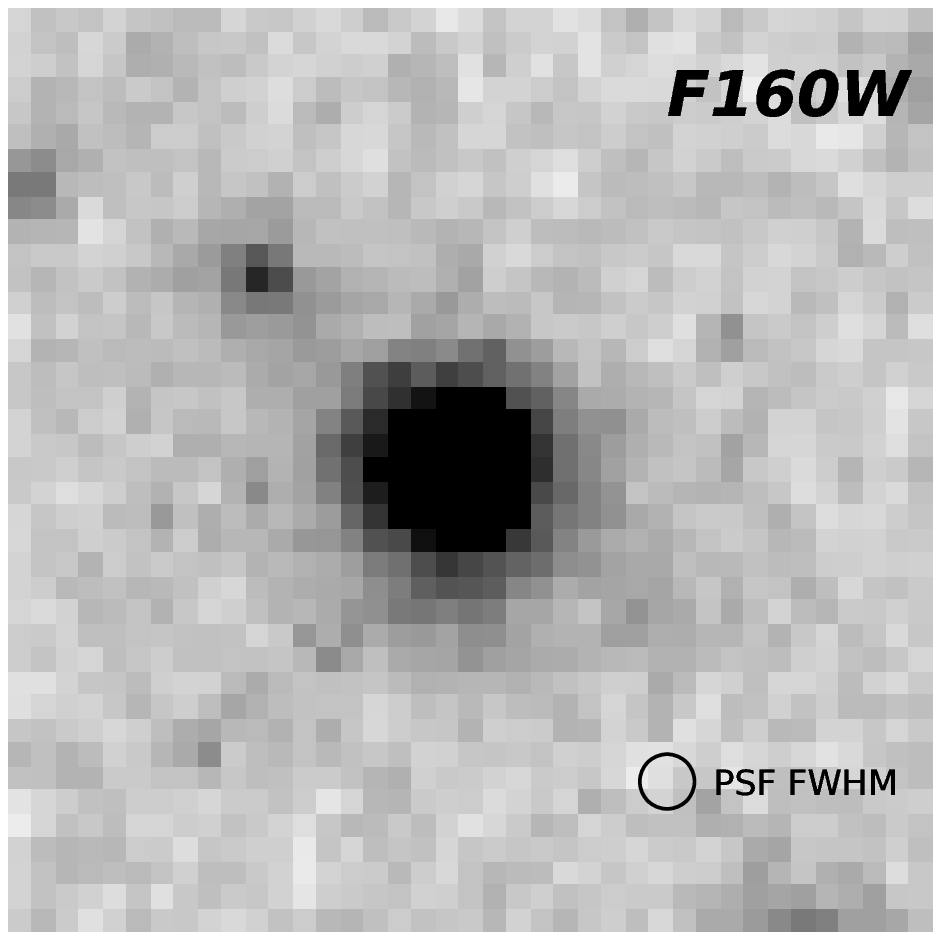}
    \includegraphics*[width=0.30\textwidth]{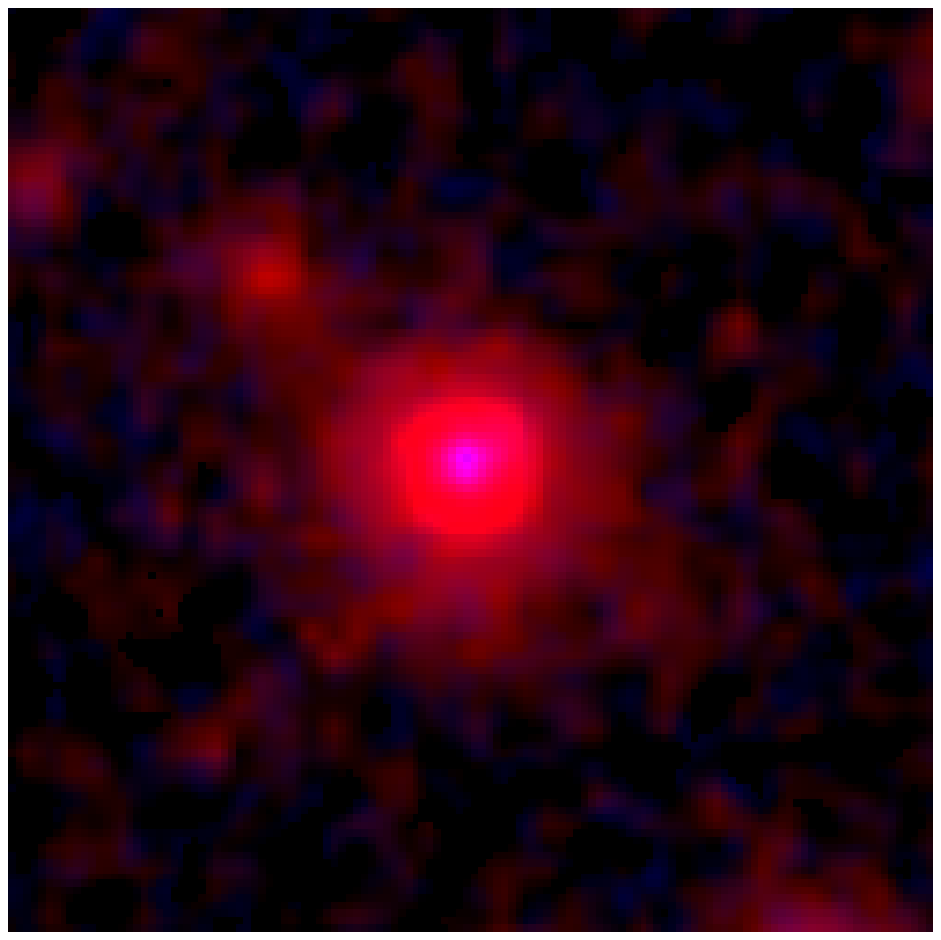}

    \includegraphics*[width=0.30\textwidth]{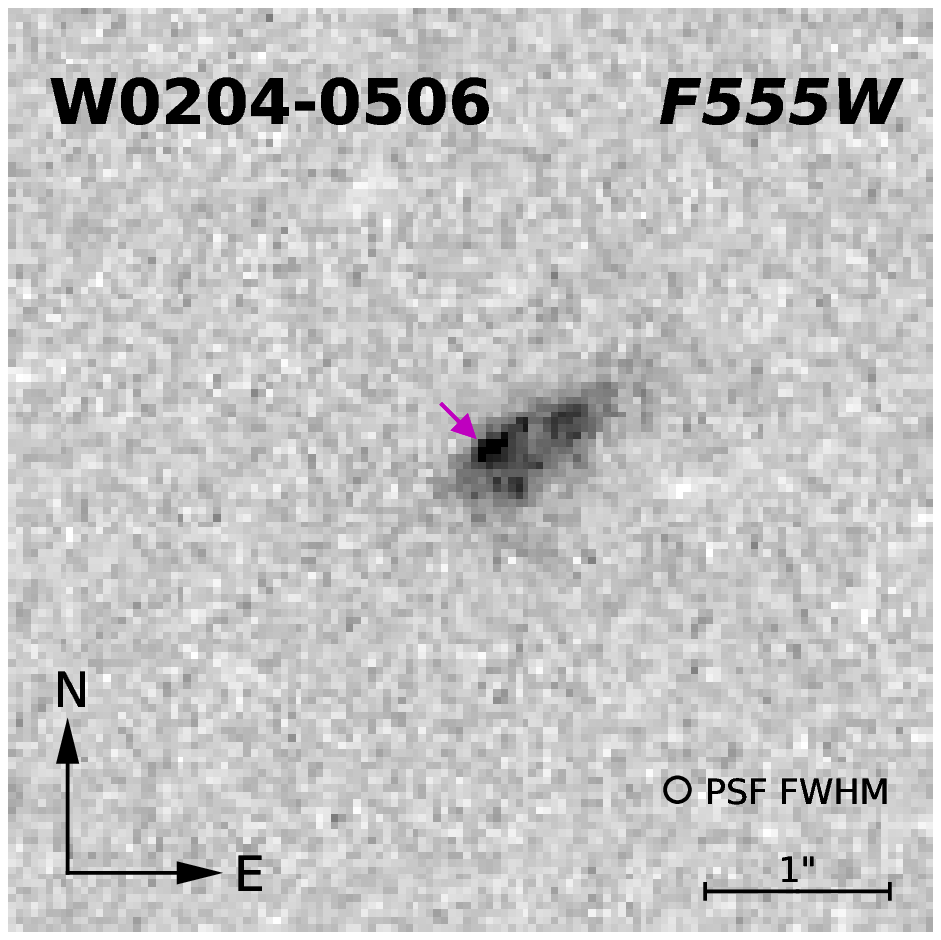}
    \includegraphics*[width=0.30\textwidth]{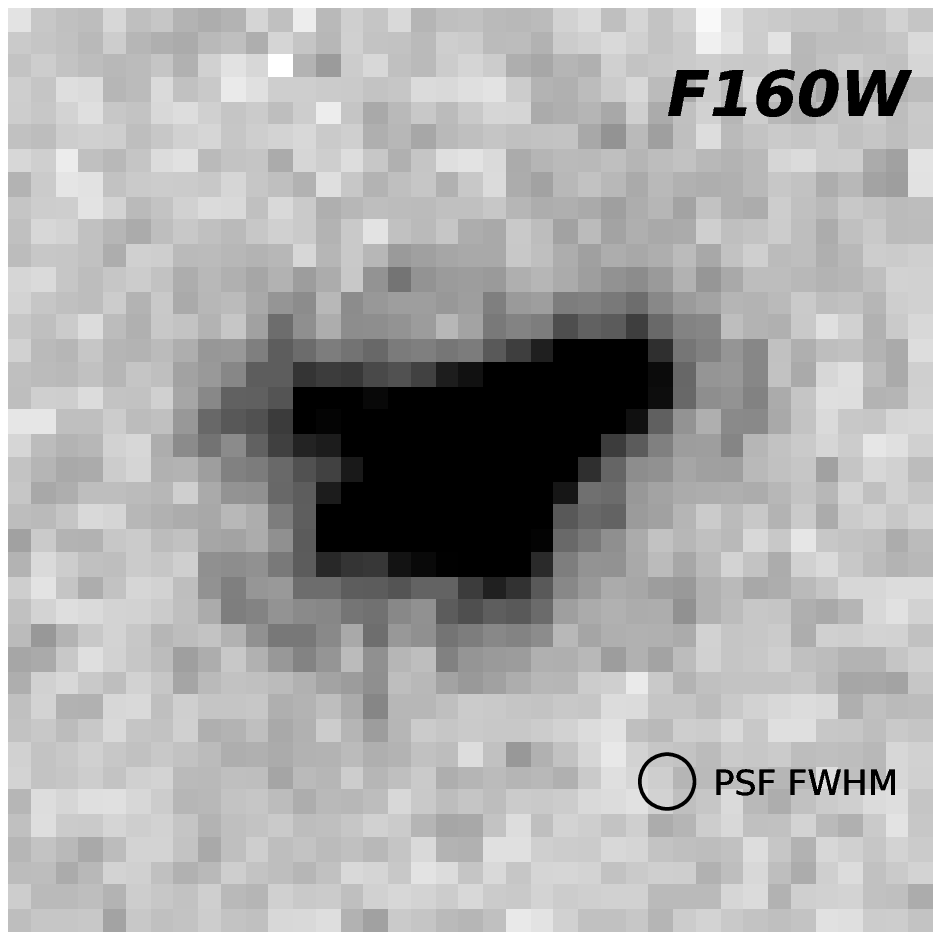}
    \includegraphics*[width=0.30\textwidth]{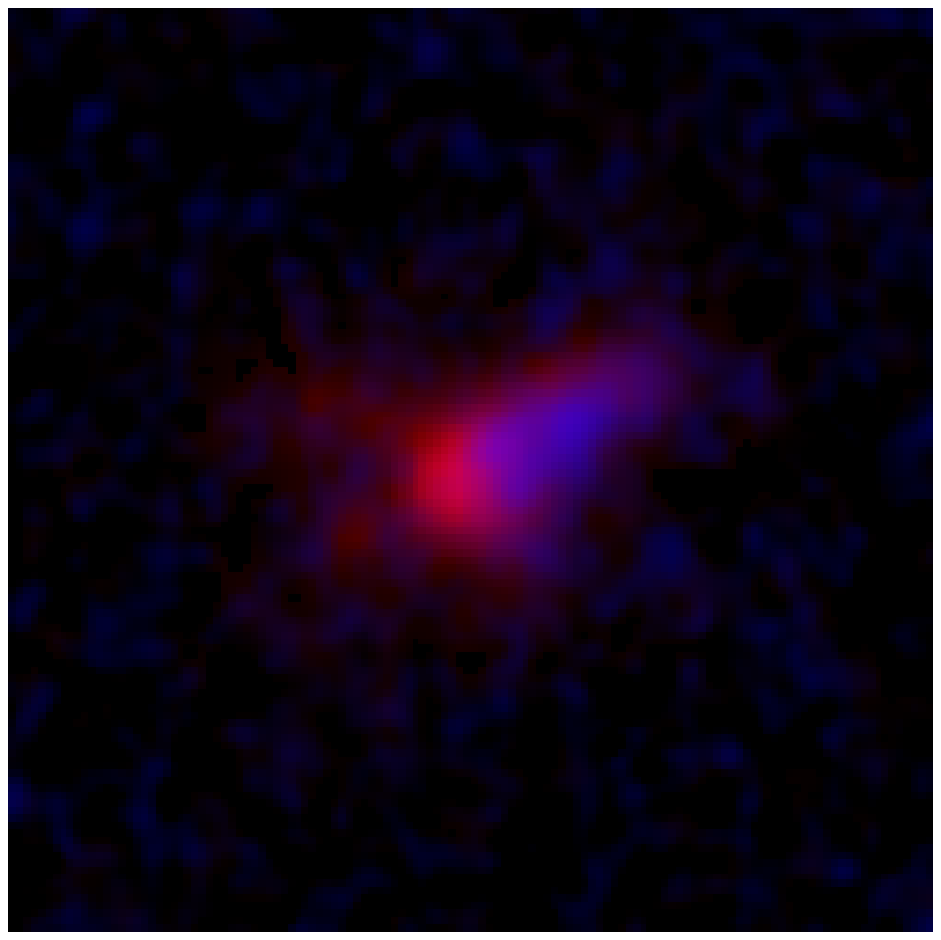}

    \includegraphics*[width=0.30\textwidth]{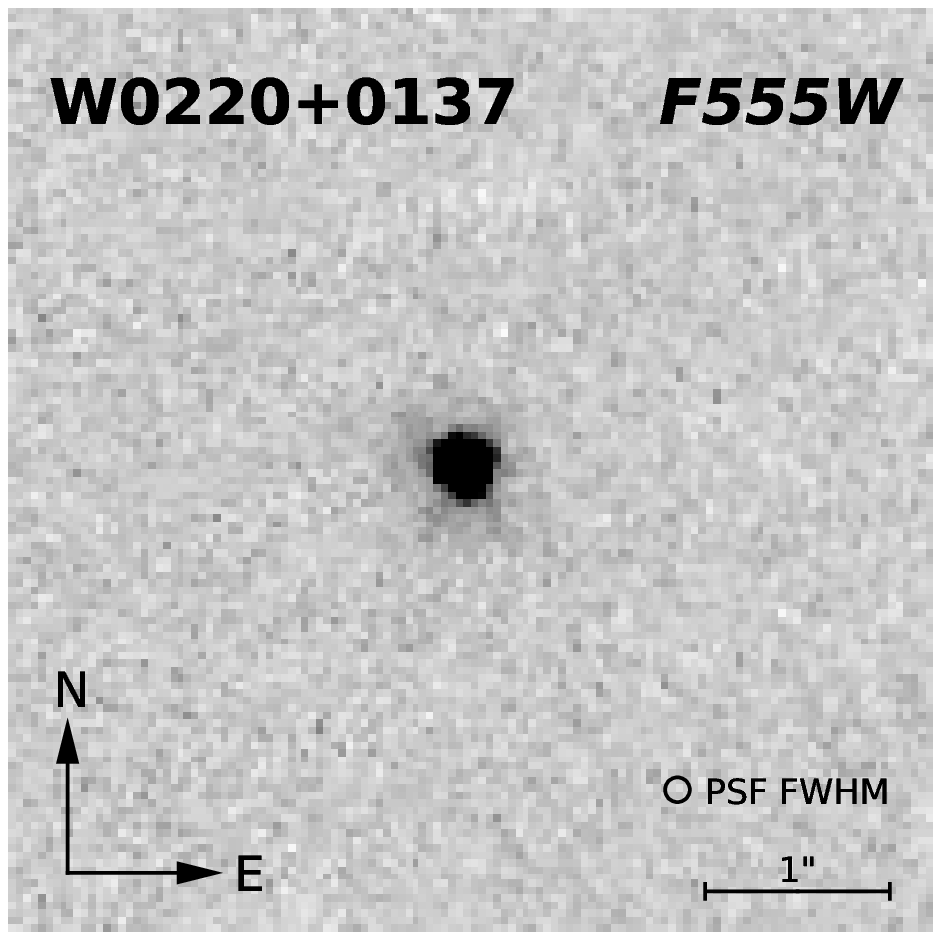}
    \includegraphics*[width=0.30\textwidth]{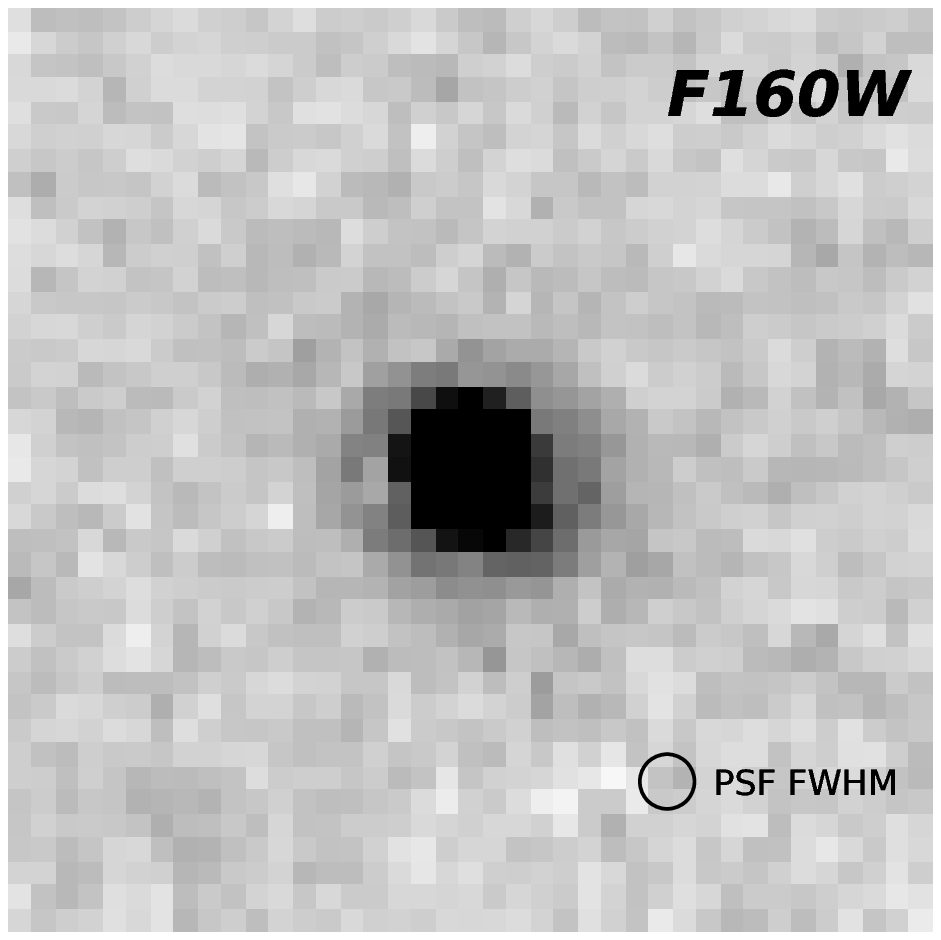}
    \includegraphics*[width=0.30\textwidth]{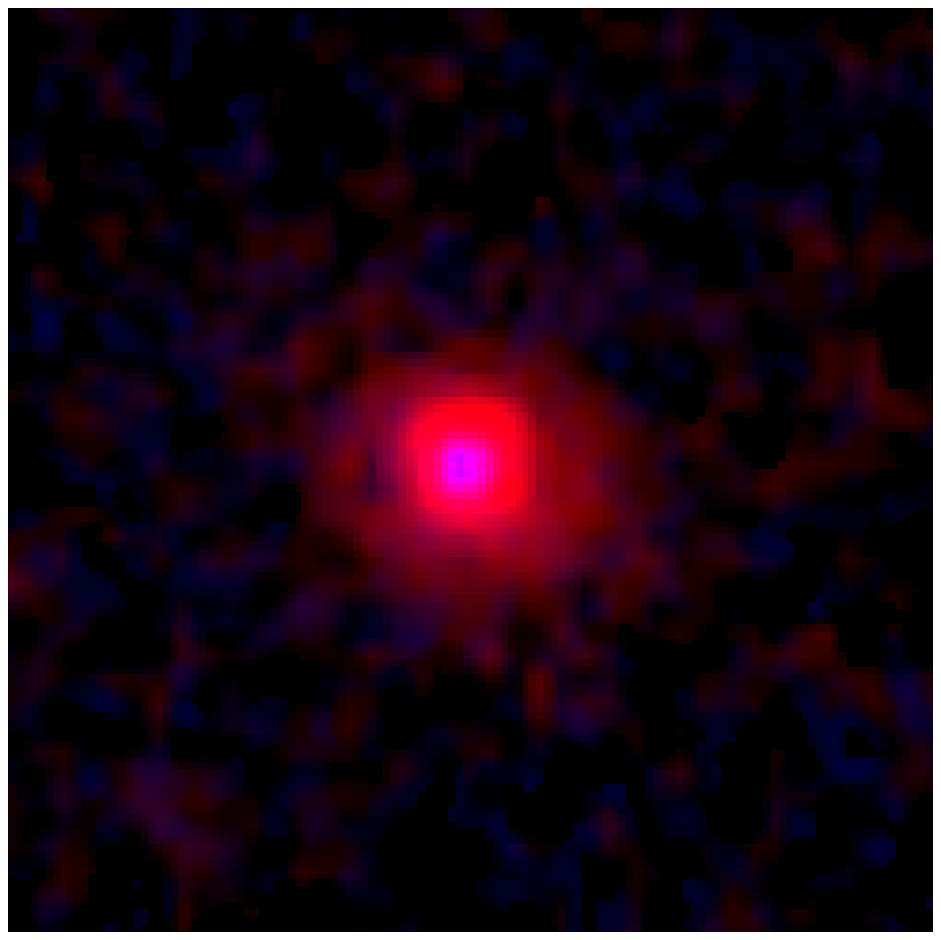}

    \caption{{\it{HST}}/WFC3 images of W0116--0505 (top row),
      W0204--0506 (middle row) and W0220+0137 (bottom row) in the
      F555W (left panels) and F160W (middle panels) bands. The right
      panels show a color-composite where the F160W band has been
      mapped to red and the F555W band has been mapped to blue, and we
      have matched the PSF of the F555W band to that of F160W. Each
      panel shows a 5\arcsec$\times$5\arcsec\ region centered on the
      F160W centroid of each target.}
    \label{fg:HST_images}
  \end{center}
\end{figure*}

Imaging observations were obtained using the WFC3 camera onboard
{\it{HST}} of all three BHD targets in both the F555W and the F160W
bands. Each target was observed during one orbit, with two exposures
in the F555W band followed by three exposures in the F160W band. The
exposure times in the F555W band were 738~s and 626~s for each image
for W0116--0505 and W0204--0506, and 735~s and 625~s each for
W0220+0137. All exposure times in the F160W band were 353~s. For the
F160W band we use the reduced images provided by the {\it{HST}}
archive. Cutouts of 5\arcsec$\times$5\arcsec\, centered on the F160W
coordinates of the target, are shown in the middle panels of Figure
\ref{fg:HST_images}.

For the F555W band we do not use the archive provided reductions, as
the pipeline cosmic ray rejection is significantly compromised by the
acquisition of only two images. Instead, we took the fully-reduced
single frames provided by the archive, including the charge transfer
efficiency correction, and used the LACOSMIC algorithm
\citep{vandokkum01} to remove cosmic rays. We then used those
cosmic-ray corrected images to continue with the pipeline processing
and combine the frames. We aligned the F555W image to the F160W image
using stars detected in both bands. The final images are shown in the
left panels of Figure \ref{fg:HST_images}. Table \ref{tab:phot}
presents the 4\arcsec\ diameter aperture magnitudes measured in each
band for each object.

The right panel of Figure 5 show an RGB composite of the images
created using the \citet{lupton04} algorithm as implemented through
the {\tt{astropy v2.0.1}}\footnote{\url{http://www.astropy.org/}}
function {\tt{make\_lupton\_rgb}}. We assigned the F555W image to the
blue channel and the F160W image to the red channel, while leaving the
green channel empty. Before producing the RGB composite, we convolve
the F555W image with a Gaussian kernel to match its PSF to that of the
F160W image. We assume that the PSFs of both images are well modeled
by Gaussian PSFs with the respective FWHM as provided by the WFC3
documentation\footnote{\url{http://www.stsci.edu/hst/wfc3/ins\_performance/ground/components/filters},\\\url{http://www.stsci.edu/hst/wfc3/documents/handbooks/currentIHB/c06\_uvis07.html\#391868}},
namely 0.067\arcsec\ for the F555W channel, and 0.148\arcsec\ for the
F160W channel. Hence, the Gaussian kernel used on the F555W image
corresponds to a Gaussian function with $\rm FWHM_{\rm kernel}^2 = \rm
FWHM_{\rm F160W}^2 - FWHM_{\rm F555W}^2$.

The emission of the three objects is clearly
resolved in both bands. For W0116--0505 and W0220+0137 the
morphologies seem to be broadly undisturbed in both bands, with the
F160W emission having a larger extent and a higher luminosity. The
emission peaks in both bands are spatially co-located. W0204--0506 is,
on the other hand, quite clearly disturbed, with the F160W morphology
(rest-frame 5200\AA) suggestive of a recent interaction. The F555W
emission (rest-frame 1800\AA) is patchy, reminiscent of a
starburst. We discuss the implications of this UV morphology further
in \S\ref{sec:discussion}.

To more quantitatively assess the morphology of these sources, we have
measured different coefficients commonly used in the
literature. Specifically, we follow \citet{lotz04} to measure the
Gini, $M_{20}$ and $A$ coefficients \citep[][and references
  therein]{lotz04}. The Gini coefficient \citep{abraham03} measures
how uniformly distributed is the light among the pixels of a galaxy in
an image, such that Gini is 0 if all pixels have a uniform brightness
and is 1 if all brightness is concentrated in a single pixel. The
$M_{20}$ coefficient measures the second order moment of the brightest
20\% of the flux of the galaxy as compared to the total second order
moment, $M_{\rm tot}$. The moments are computed around a center chosen
to minimize $M_{\rm tot}$. The $A$ coefficient measures the rotational
asymmetry of a galaxy by subtracting an image of the galaxy rotated by
180 degrees. The rotational center is chosen to minimize $A$. For
further details on these coefficients, we refer the reader to
\citet{lotz04} and \citet{conselice14}.

We start by subtracting the background using {\tt{SExtractor}}
\citep[{\tt{v2.19.5}},][]{bertin96} as well as obtaining the centroid
of each object in each band. We then compute the Petrosian radius
\citep{petrosian76} and generate the segmentation map following
\citet{lotz04}, and finally proceed to measure the coefficients
discussed above. The values and uncertainties of the Gini, $M_{20}$
and $A$ coefficients for each object in each band are shown in Table
\ref{tab:morph}. We estimate the uncertainties in each parameter
through a Monte Carlo approach. For a given object in a given band, we
use the uncertainty in each pixel to generate 1,000 resampled images
assuming Gaussian statistics. We then repeat the measurement in each
resampled image following the procedure outlined above. We assign the
measurement error to be the dispersion of the coefficient measurements
in the 1,000 resampled images.

\begin{deluxetable*}{l c c c c}

  \tablecaption{Morphology of Blue Excess Hot DOGs\label{tab:morph}}

  \tablehead{
    \colhead{Source} &
    \colhead{Band} &
    \colhead{Gini} &
    \colhead{$M_{20}$} &
    \colhead{$A$}
  }

  \tabletypesize{\small}
  \tablewidth{0pt}
  \tablecolumns{5}

  \startdata
  W0116--0505 & F555W & 0.499$\pm$0.003 & --2.00$\pm$0.16 & 0.133$\pm$0.010\\
              & F160W & 0.527$\pm$0.008 & --2.08$\pm$0.02 & 0.116$\pm$0.049\\
  W0204--0506 & F555W & 0.529$\pm$0.016 & --1.19$\pm$0.14 & 0.599$\pm$0.027\\
              & F160W & 0.633$\pm$0.011 & --0.81$\pm$0.02 & 0.278$\pm$0.010\\
  W0220+0137  & F555W & 0.491$\pm$0.004 & --1.77$\pm$0.09 & 0.112$\pm$0.009\\
              & F160W & 0.559$\pm$0.016 & --2.16$\pm$0.04 & 0.172$\pm$0.016\\
  \enddata

\end{deluxetable*}

Recently, \citet{farrah17} measured these coefficients for 12 Hot DOGs
using {\it{HST}}/WFC3 images in the F160W. Using the boundaries
proposed by \citet{lotz04} in the Gini--$A$ plane and by
\citet{lotz08} in the Gini--$M_{20}$ plane, \citet{farrah17}
determined that while Hot DOGs have a high merger fraction ($\sim
80\%$), this fraction is consistent with that found for massive
galaxies at $z\sim 2$, leading them to conclude that Hot DOGs are not
preferentially associated with mergers. These results are generally
consistent with those of \citet{fan16b} who also found a high merger
fraction ($62\pm 14\%$) among Hot DOGs, as well as with those recently
presented by \citet{diaz18}, who found evidence with sub-mm ALMA $\sim
200~\mu\rm m$ imaging of a triple major merger in the the most
luminous Hot DOG, W2246--0526. If we adopt the same boundaries used by
\citet{farrah17} to classify our sources according to their F160W
morphologies, and noting that all caveats identified by
\citet{farrah17} also apply here, we find that the host galaxies of
W0116--0505 and W0220+0137 are not consistent with mergers but instead
are classified as undisturbed early-type galaxies. For W0204--0506, on
the other hand, we find that its host galaxy morphology is best
classified as an on-going merger. These results are consistent with
our visual characterization of the host galaxies.

\subsection{Chandra Observations}\label{ssec:xray_data}

We have obtained {\it{Chandra}}/ACIS-S observations of two of our
targets: W0116--0505 and W0220+0137 (proposal ID 17700696). Each
object was observed with a total exposure time of 70~ks. W0116--0505
was observed continuously, while the observations of W0220+0137 were
split into one 30~ks and two 20~ks visits spread throughout seven
days. It is worth noting that these observations have previously been
presented by \citet{vito18} in the context of a larger sample of Hot
DOGs observed in X-rays. They find both sources are heavily absorbed
at those wavelengths. \citet{goulding18} analyzed the observations for
W0220+0137 as well, but in the context of a large sample of Extremely
Red Quasars (ERQs), and also found the source to be heavily absorbed
at X-ray energies, qualitatively consistent with the rest of the ERQ
population analyzed. Here we analyze the data following the approach
of A16, who analyzed the archival {\it{Chandra}}/ACIS-I observations
of W0204--0506.

We use {\sc{ciao}} v4.7 to analyze these data. The spectral data
products, including the source and background spectra, and the
response files were created using the {\tt{specextract}} tool. Source
events were extracted from circular regions with 2\arcsec\ radii
centered on the source, while background events were extracted from
annuli with inner and outer radii of 3 and 6\arcsec, respectively. For
W0220+0137, the spectral products from the three observations were
combined into one using the tool {\tt{combine\_spectra}}. We confirm
that combining the spectral products in this way does not
systematically affect our results by fitting the spectra of each
individual obsID simultaneously, linking the fit parameters between
them. We find consistent results compared to fitting the combined
spectral products and proceed with the combined products. We use the
{\sc{heasoft}} tool {\tt{grppha}} to group the spectra with a minimum
of one count per bin.

\begin{figure}
  \begin{center}
    \plotone{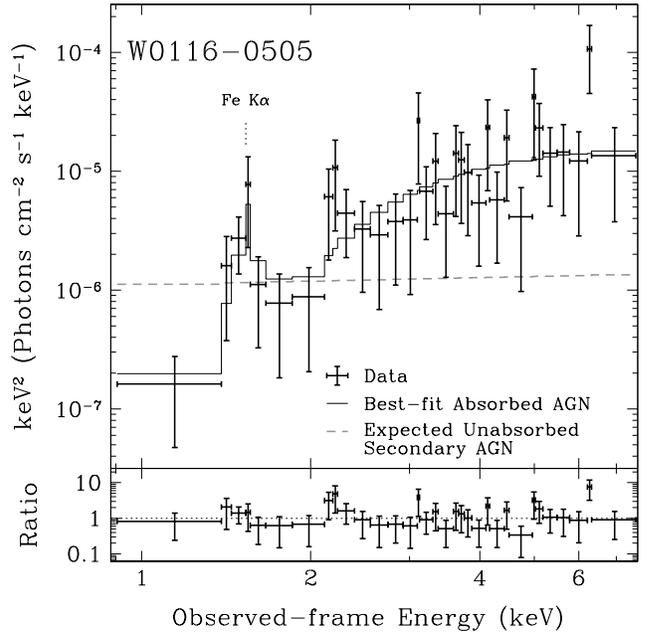}
    \caption{{\it{(Top panel)}} Unfolded X-ray spectrum of W0116--0505
      obtained using {\it{Chandra}}/ACIS-S (see \S\ref{ssec:xray_data}
      for details). The solid black line shows the best-fit absorbed
      AGN model to the spectrum, as described in
      \S\ref{sec:xray_model}. The dashed-gray line shows the emission
      expected for a second, unobscured AGN in the system powering the
      observed UV/optical emission. {\it{(Bottom panel)}} The data
      points show the ratio between the observed spectrum and the
      best-fit model.}
    \label{fg:W0116_xray_spec}
  \end{center}
\end{figure}

\begin{figure}
  \begin{center}
    \plotone{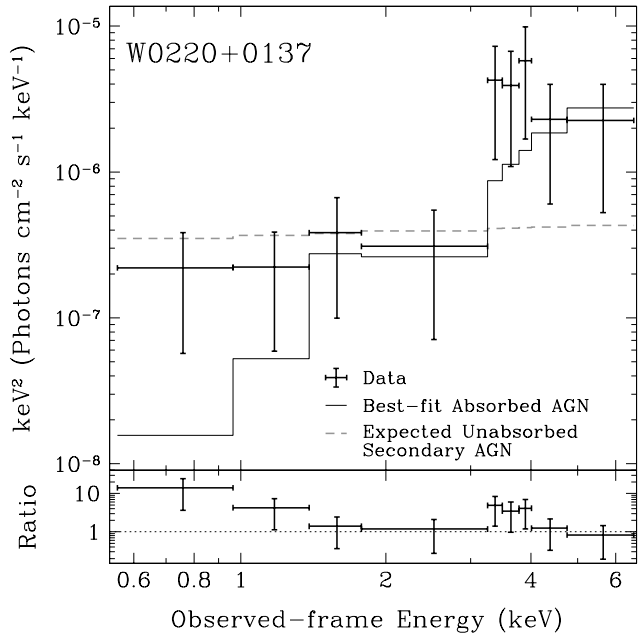}
    \caption{Same as Fig. \ref{fg:W0116_xray_spec} but for the
      {\it{Chandra}}/ACIS-S spectrum of W0220+0137.}
    \label{fg:W0220_xray_spec}
  \end{center}
\end{figure}

\begin{figure}
  \begin{center}
    \plotone{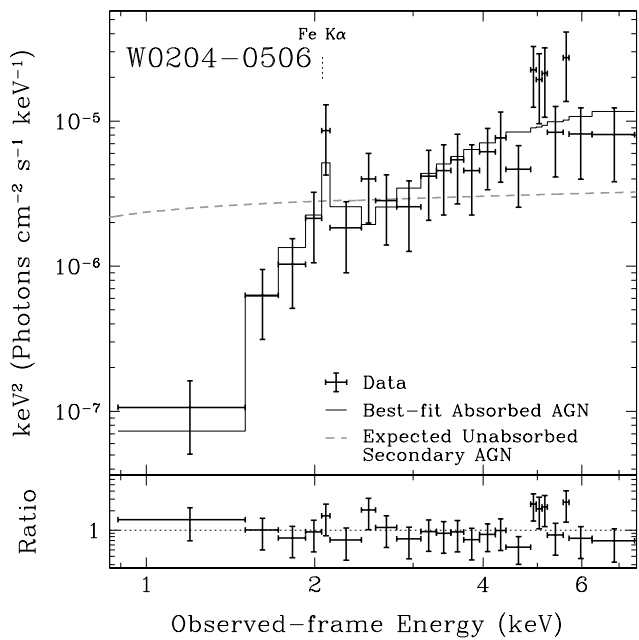}
    \caption{Same as Fig. \ref{fg:W0116_xray_spec} but for the
      {\it{Chandra}}/ACIS-I spectrum of W0204--0506. Adapted from
      Fig. 4 of A16.}
    \label{fg:W0204_xray_spec}
  \end{center}
\end{figure}

After subtracting the background, 74 counts are detected for
W0116--0505, and only 18 for W0220+0137. Figures
\ref{fg:W0116_xray_spec} and \ref{fg:W0220_xray_spec} show their
respective unfolded spectra. Note that the spectra have only been
unfolded for the benefit of their presentation. For reference, we also
show the ACIS-I spectrum of W0204--0506 in Figure
\ref{fg:W0204_xray_spec}, which had a significantly longer exposure
time of 160~ks. The shape of all three spectra differ significantly
from that of an unabsorbed power-law, suggesting the emission is
dominated by a highly obscured AGN, as expected from the SED modeling
presented in \S\ref{ssec:bhds}. In the next section we model these
spectra and discuss their implications for the nature of the BHDs.

\section{X-ray Data Modeling}\label{sec:xray_model}

The X-ray spectra of W0116--0505 and W0220+0137 are clearly hard,
implying the emission is most likely dominated by a highly obscured
AGN. To better constrain the properties of the obscured AGN, we fit
the emission of both objects using the models of \citet{brightman11},
following the same approach as in A16. These models predict the X-ray
spectrum as observed through an optically thick medium with a toroidal
geometry, as posited by the AGN unified scheme. The models employ
Monte-Carlo techniques to simulate the transfer of X-ray photons
through the optically-thick neutral medium, self-consistently
including the effects of photoelectric absorption, Compton scattering
and fluorescence from Fe K, amongst other elements. Treating these
effects self consistently rather than separately has the advantage of
reducing the number of free parameters and of gaining constraints on
the spectral parameters. It is therefore particularly useful for low
count spectra such as those we are fitting here. We therefore carry
out the parameter estimation by minimizing the Cash statistic
\citep{cash79}, modified through the W-statistic provided by
XSPEC\footnote{\url{https://heasarc.gsfc.nasa.gov/xanadu/xspec/manual/XSappendixStatistics.html}}
to account for the subtracted background. Also following the approach
of A16, we require the photon index $\Gamma$ to be $\ge 1.6$, as it is
poorly constrained by our data and lower values are only appropriate
for low Eddington ratios. In practice, the fitting procedure we used
allows values of $\Gamma$ in the range 1.6--3.0, and values of $N_{\rm
  H}$ in the range $10^{20}-10^{26}~\rm cm^{-2}$. However, for
W0220+0137, there are too few counts to constrain all spectral
parameters, therefore we fix $\Gamma$ to the canonical value of 1.9,
leaving only $N_{\rm H}$ and the normalization free.

\begin{figure}
  \begin{center}
    \plotone{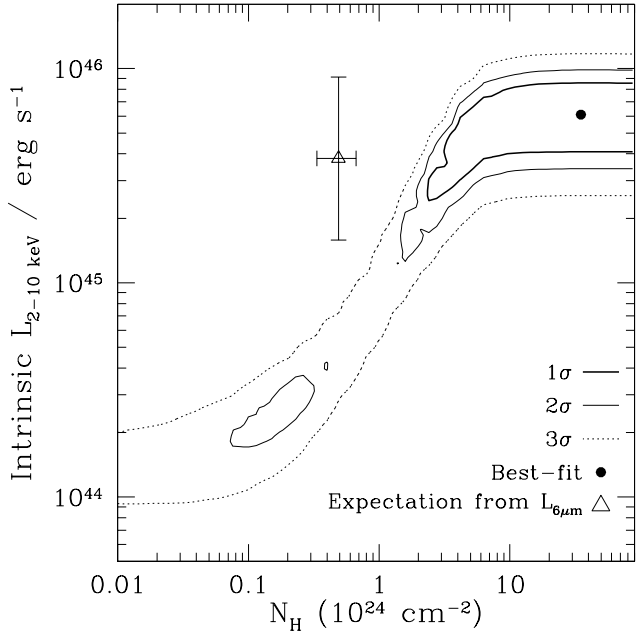}
    \caption{Confidence intervals of the fit to the
      {\it{Chandra}}/ACIS-S spectrum of W0220+0137 as described in the
      text.}
    \label{fg:W0220_contours}
  \end{center}
\end{figure}

Figures \ref{fg:W0116_xray_spec} and \ref{fg:W0220_xray_spec} show the
best-fit models to the spectra of W0116--0505 and W0220+0137,
respectively. The best-fit absorbed AGN model to W0116--0505 has an
absorption column density of neutral hydrogen of $N_{\rm H} =
1.2^{+1.0}_{-0.7}\times 10^{24}~\rm cm^{-2}$, a photon-index of
$\Gamma = 1.9^{+0.7}_{-l}$ (where $-l$ signifies that the parameter is
bound by the minimum value allowed by the fitting procedure) and an
absorption-corrected luminosity of $\log L_{2-10~\rm keV}/\rm erg~\rm
s^{-1} = 45.63^{+0.58}_{-0.24}$. The best-fit model has a Cash
statistic of $C=56.3$ for $\nu=65$ degrees of freedom. For W0220+0137,
the best-fit model has $N_{\rm H} = 3.5^{+u}_{3.2}\times 10^{25}~\rm
cm^{-2}$ (where $+u$ signifies that the parameter is bound by the
maximum value allowed by the fitting procedure) and $\log L_{2-10~\rm
  keV}/\rm erg~\rm s^{-1} = 45.79^{+0.12}_{-0.49}$, with $C=26.34$ and
$\nu=16$. Given the low number of counts for W0220+0137, Figure
\ref{fg:W0220_contours} shows the confidence regions for the two
parameters to highlight the degeneracies between them. The best-fit
values of $N_{\rm H}$ and $L_{2-10~\rm keV}$ are consistent within the
uncertainties with those found by \citet{vito18} for both sources. The
best-fit values for W0220+0137 are also consistent within the (large)
error bars with those found by \citet{goulding18}. For W0204--0506,
A16 found that the best-fit absorbed AGN has $N_{\rm H} =
0.63^{+0.81}_{-0.21}\times 10^{24}~\rm cm^{-2}$, $\Gamma =
1.6^{+0.8}_{-0.0}$ and $\log L_{2-10~\rm keV}/\rm erg~\rm s^{-1} =
44.9^{+0.86}_{-0.14}$, with $C=66.08$ and $\nu=77$.

The spectra of all three objects are likely dominated by a luminous AGN with
very high absorption. In the case of W0116--0505 and W0220+0137, the absorption
is consistent with the objects being Compton-thick (i.e., $N_{ rm H} > 1.5\times
10^{24}~\rm cm^{-2}$). This is in qualitative agreement with the SED modeling
presented in \S\ref{ssec:bhds}. From the SED model of each object we can
estimate the rest-frame intrinsic (i.e., obscuration corrected) specific
luminosity at 6$\mu$m, $L_{6\mu\rm m}$, which has been shown to be well
correlated with the $L_{2-10~\rm keV}$ X-ray luminosity by a number of authors
\citep{fiore09,ghandi09,bauer10,mateos15,stern15,chen17}. We use the best-fit
relation of \citet{stern15} between $L_{6\mu\rm m}$ and $L_{2-10~\rm keV}$,
which accurately traces this relation up to very high $L_{6\mu\rm m}$ and is
hence most appropriate for our targets. From the $L_{6\mu\rm m}$ of the most
luminous and obscured AGN component of W0116--0505, this relation predicts $\log
L^{Predicted}_{2-10~\rm keV}/\rm erg~\rm s^{-1} = 45.53\pm 0.62$, which is in
excellent agreement with the luminosity of the best-fit model to the X-ray data
of $\log L_{2-10~\rm keV}/\rm erg~\rm s^{-1} = 45.63^{+0.58}_{-0.24}$. For
W0220+0137 we also find excellent agreement, with $\log L^{Predicted}_{2-10~\rm
keV}/\rm erg~\rm s^{-1} = 45.58\pm 0.62$ and $\log L_{2-10~\rm keV}/\rm erg~\rm
s^{-1} = 45.54^{+0.30}_{-2.16}$. A16 nominally found a good agreement as well
for W0204--0506, as they estimated $\log L^{Predicted}_{2-10~\rm keV}/\rm
erg~\rm s^{-1} = 45.36\pm 0.37$ and found $\log L_{2-10~\rm keV}/\rm erg~\rm
s^{-1} = 44.9^{+0.86}_{-0.14}$ from the best-fit X-ray model. However, when
jointly considering this with the best-fit and expected absorption, their Figure
5 suggests W0204--0506 may be somewhat X-ray weak. We remind the reader,
nonetheless, that the error bars derived for our $L_{6\mu\rm m}$ estimates could
be underestimated when considering the minimalist approach used for the SED
modeling in \S\ref{ssec:bhds}, which may make the disagreement less
significant.

From the SED modeling we also have an estimate of the amount of dust
that is obscuring the luminous AGN that dominates in both the mid-IR
and the X-rays. Comparing to the column densities of neutral hydrogen
constrained by the modeling of X-ray spectra, we find dust-to-gas
ratios of $E(B-V)/N_{\rm H} = 3.5\pm 2.0 \times 10^{-24}~\rm cm^2~\rm
mag$ for W0116--0505, where the uncertainty corresponds to the 68.3\%
confidence interval and has been derived, for simplicity, assuming
Gaussian statistics. For W0220+0137 we find $E(B-V)/N_{\rm H} = 2.3
\times 10^{-24}~\rm cm^2~\rm mag$. As $N_{\rm H}$ is not constrained
at the 90\% level within the model boundaries, we cannot derive a
meaningful confidence interval. For W0204--0506, A16 found a larger
ratio of $E(B-V)/N_{\rm H} = 1.54\pm 1.26 \times 10^{-23}~\rm cm^2~\rm
mag$. For comparison, the median dust-to-gas ratio in AGN found by
\citet{maiolino01} is $1.5\times 10^{-23}~\rm cm^2~\rm mag$. This
value is comparable to that found in W0204--0506, while those found in
W0116--0505 and W0220+0137 are lower. Unfortunately the large
uncertainties in this quantity make this result difficult to
interpret, but it is worth noting that recently \citet{yan19}
identified a very low dust-to-gas ratio of $\approx 4\times
10^{-25}\rm cm^{2}$ for a heavily obscured nearby quasar at $z=0.218$
with $N_{\rm H}\approx 3\times 10^{25}~\rm cm^{-2}$, with around
$N_{\rm H}\approx 10^{23}~\rm cm^{-2}$ coming from the ISM, which
could be a better analog to our objects. If the dust-to-gas ratio is
indeed significantly lower in W0116--0505 and W0220+0137 than in
W0204--0506, it could either imply a low metallicity for the former
systems such that there is a deficit of dust overall in the host
galaxy, or that a higher than typical fraction of absorbing gas exists
within the dust sublimation radius of the accretion disk. We speculate
the latter could be consistent with the recent results of \citet{wu18}
that show Hot DOGs are accreting close to the Eddington limit, perhaps
as a result of higher gas densities in the vicinity of the SMBH.

Taken together, these results could imply that W0116--0505 and
W0220+0137 represent a different class of object than W0204--0506, as
the former are either dust-poor or gas-rich in the nuclear regions,
but have normal X-ray luminosities, while the latter has a normal
amount of dust but might be somewhat X-ray weak. The morphology of the
{\it HST} imaging strongly differs between these objects, as discussed
in \S\ref{ssec:hst_data}, further supporting this. \citet{goulding18}
points out that W0220+0137 is also classified as an ERQ by
\citet{hamann17}, and finds that W0116--0505 fulfills most of the
criteria and hence classifies it as ERQ-like. This supports a view in
which ERQs and Hot DOGs are not independent populations, but possibly
related to each other with BHDs being the link between them. We
speculate that Hot DOGs might correspond to the highly obscured AGN
phase of galaxy evolution proposed by, e.g., \citet{hopkins08} or
\citet{alexander12}, and as the obscuration starts clearing out
\citep[see][for a description of the different physical scales of the
  obscuring materials]{hickox18}, the object transforms into a BHD and
then an ERQ, before transitioning into an unobscured quasar. The
significant levels of outflowing ionized gas identified by
\citet{zakamska16} for four ERQs, by \citet{diaz16} for the most
luminous Hot DOG, W2246--0526, and by \citet{wu18} for two more Hot
DOGs, support the view that both types of objects are experiencing
strong AGN feedback.

\section{Source of the Excess Blue Emission}\label{sec:discussion}

\subsection{Dual AGN}

One of the possible scenarios proposed by A16 is that BHDs could be powered by
two AGNs instead of one, where a primary luminous, highly obscured AGN dominates
the mid-IR emission, and a secondary fainter, unobscured or lightly obscured AGN
dominates the UV/optical emission. As can be seen in Table \ref{tab:pars}, the
putative secondary AGN has a much lower ($\lesssim 1\%$) best-fit monochromatic
luminosity at 6$\mu$m in all three targets. As discussed above, the former would
be expected to dominate the hard X-ray emission of these sources, and that is
exactly what is observed. However, the less luminous component would contribute
significant soft X-ray emission, that can be constrained by the {\it{Chandra}}
observations. In Table \ref{tab:pars} we list the expected intrinsic 6$\mu$m
luminosity of the primary and secondary best-fit AGN components for both
W0116--0505 and W0220+0137. It is important to note that for the secondary AGN
components we have no useful constraints in the IR, as the rest-frame near-IR is
dominated by the host galaxy and the mid-IR is dominated by the primary AGN
component. Its 6$\mu$m luminosity comes instead indirectly from the template fit
to the rest-frame UV/optical SED. Furthermore, there could be neglected
systematic uncertainties, as discussed at the end of \S\ref{ssec:bhds}. As we
did in \S\ref{sec:xray_model}, we can estimate the expected 2--10~keV luminosity
using the relation of \citet{stern15}. Hence, if the secondary component is a
real second AGN in the system, for W0116--0505 we expect it to have an X-ray
luminosity of $\log L_{2-10~\rm keV}^{Predicted}/\rm erg~\rm s^{-1} = 44.43\pm
0.37$, and for W0220+0137 we expect it to have $\log L_{2-10~\rm
keV}^{Predicted}/\rm erg~\rm s^{-1} = 44.29\pm 0.62$. The gray-dashed curves in
Figures \ref{fg:W0116_xray_spec} and \ref{fg:W0220_xray_spec} show the expected
X-ray spectrum of these secondary components for W0116--0505 and W0220+0137
respectively. We assume power-law spectra with $\Gamma=1.9$ and no absorption,
as both secondary components show no reddening in the UV/optical. Figure
\ref{fg:W0204_xray_spec} also shows the expected X-ray spectrum of the secondary
AGN, as determined from the analysis of A16.

If we add a secondary power-law component to the X-ray spectral fit,
we can place 90\% upper limits on the luminosity of these power-law
components of $\log L_{2-10~\rm keV}/\rm erg~\rm s^{-1} < 43.95$ in
W0116--0505, and of $\log L_{2-10~\rm keV}/\rm erg~\rm s^{-1} < 43.93$
in W0220+0137. These limits are marginally consistent with the
2--10~keV luminosities expected given the optical/UV luminosities
observed. For W0204--0506 on the other hand, A16 was able to rule out
this scenario with high confidence. Unlike the analysis presented
here, A16 reached this conclusion by comparing the change in Cash
statistic of the X-ray spectra modeling obtained by requiring or not
the presence of the secondary AGN emission with the expected
luminosity. Specifically, A16 found that including the secondary
component resulted in an increase in the Cash statistic $\Delta
C=128.38$, which allowed to rule out the dual AGN scenario with
$>99.9\%$ confidence. We do not replicate this analysis for
W0116--0505 and W0220+0137, as the interpretation of the change in the
$C$ statistic ($\Delta C=17.3$ and $\Delta C=18.7$, respectively) is
complicated by the lower number of counts detected, particularly in
the case of W0220+0137, and hence we cannot assume Gaussian
statistics.

Hence, the X-ray spectra of all three objects are better described by
the single, highly absorbed AGN model, suggesting that BHDs are not
dual AGN. The case is strongest for W0204--0506, while for W0116--0505
and W0220+0137 we cannot completely reject the dual AGN scenario with
high confidence using the current data sets.

\subsection{Extreme Star-formation}

Another possibility discussed by A16 is that the
UV/optical SED of BHDs is powered by unobscured extreme star-formation
rather than by unobscured AGN emission. This would account for a very
blue UV/optical SED without the X-ray contribution expected for a
secondary AGN.

This scenario was studied in detail by A16 for W0204--0506. Modeling
the UV/optical SED of this object using the Starburst99 v7.0.0 code
\citep{leitherer99,leitherer10,leitherer14,vazquez05} in combination
with the EzGal package of \citet{mancone12}, they determined that the
SED could be consistent with being powered by a young starburst but
only if the SFR was very high. Specifically, they assumed the latest
Geneva models available for the used version of Starburst 99
\citep[see][for details]{leitherer14}, a constant SFR, and a solar
metallicity, and determined that the SED could only be powered by a
starburst of age $\lesssim$5~Myr with ${\rm SFR}\gtrsim
1000~M_{\odot}~\rm yr^{-1}$ with 90\% confidence. A lower metallicity
somewhat eases these constraints, with the lowest metallicity
available for the Geneva models in Starburst99 of $Z=0.001$ implying
$\rm age\lesssim 100~\rm Myr$ and ${\rm SFR}\gtrsim 250~M_{\odot}~\rm
yr^{-1}$. However, A16 considered that such a low metallicity was
unlikely given the large amount of dust available in the inner regions
of the system that give rise to the high specific luminosities in the
mid-IR. Furthermore, due to the large, unobscured SFR implied by the
solar metallicity models, A16 considered that the UV/optical SED was
unlikely powered by a starburst.

However, the morphology of the UV emission in the {\it{HST}} imaging
we have obtained (see \S\ref{ssec:hst_data} for details) seems to
imply that starburst activity is present in W0204--0506. As shown in
Figure \ref{fg:HST_images}, the flux traced by the F555W band
(rest-frame $\sim$1750\AA) is distributed along the NE section of the
galaxy, and concentrated in a few distinct regions. The bulk of the
F555W emission is considerably offset from the emission of the older
stars traced by the F160W band (rest-frame
$\sim$5100\AA). Furthermore, the morphology of the system is
consistent with a merger (see \S\ref{ssec:hst_data}), which can
trigger significant star-formation activity.

The analysis of A16 in conjunction with the {\it{HST}} imaging
available for W0204--0506 then imply that if its UV/optical SED is
solely powered by a starburst, then the system must be in a very
uncommon state. On one hand, it could be that the system has a very
large metallicity gradient, such that in the outskirts, where
star-formation dominates, the metallicity is close to primordial and
SFR is only $\gtrsim 250~M_{\odot}~\rm yr^{-1}$, yet near the SMBH the
metallicity is high enough to allow for the substantial amount of dust
needed to obscure the hyper-luminous AGN. The other possibility would
be that W0204--0506 does not have a substantial metallicity gradient
but is instead powered by the strongest unobscured starburst known
with $\rm SFR\gtrsim 1000~M_{\odot}~\rm yr^{-1}$.

A third and more likely option is that while a moderate starburst is
ongoing in the system, the UV/optical emission is still dominated by
light leaking from the central highly obscured AGN. As shown in Table
\ref{tab:morph} (also see discussion in \S\ref{ssec:hst_data}), the
light distribution of W0204--0506 in the F555W band has a somewhat
larger Gini and a significantly larger $M_{20}$ coefficient than the
other two BHDs studied. While the large $M_{20}$ is consistent with
the observed patchiness of the system, the high Gini coefficient
implies that the light is strongly concentrated in the brightest
regions. In the left panel of Figure \ref{fg:HST_images} it can be
appreciated that the NW UV clump (marked by the magenta circle,
0.2\arcsec\ diameter) is brighter than the rest, containing
approximately 10\% of the total F555W flux measured in the
4\arcsec\ radius aperture. This region is close to the geometrical
center of the F160W light distribution, and could correspond to the
position of the buried AGN. That the optical spectrum of this source
(Fig.  \ref{fg:W0204_spec}) shows a mixture of narrow and broad
emission lines is also consistent with this picture, as A16 reported a
FWHM of $1630\pm 220~\rm km~\rm s^{-1}$ for C\,{\sc iv} but of only
$550\pm 100~\rm km~\rm s^{-1}$ for $\rm{C}\,{\sc iii}]$.

\begin{figure}
  \begin{center}
    \plotone{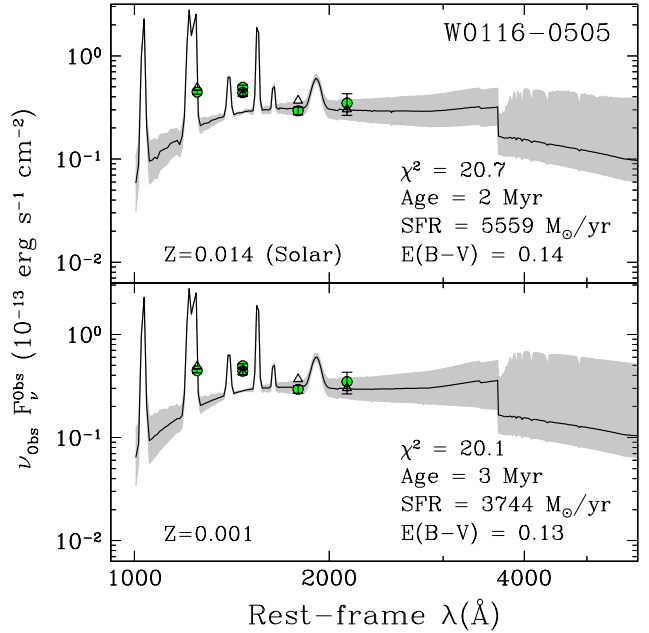}
    \caption{{\it{(Upper panel)}} The solid black line shows the
      best-fit Starburst99 SED model to the UV/optical broad-band
      photometry of W0116--0505, assuming solar metallicity (see text
      for details). The gray shaded area shows all SED shapes within
      the 90\% confidence interval. {\it{(Bottom panel)}} Same as in
      the top panel but for a metallicity of $Z=0.001$.}
    \label{fg:SF_W0116}
  \end{center}
\end{figure}

\begin{figure}
  \begin{center}
    \plotone{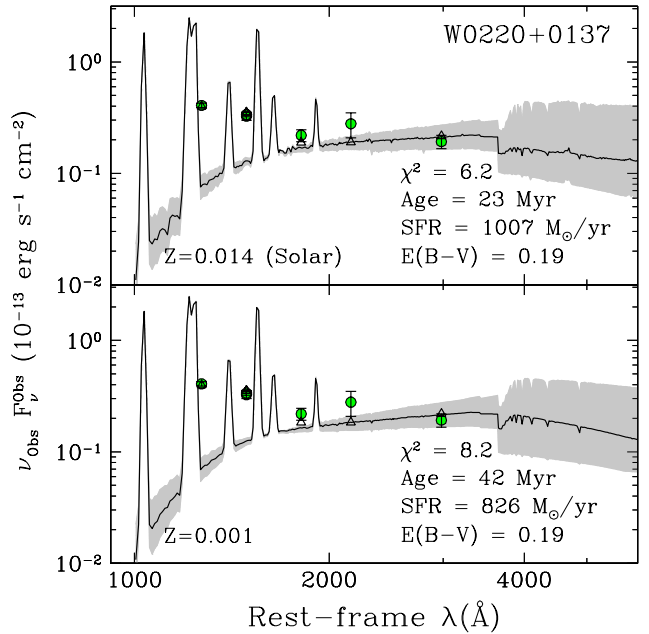}
    \caption{Same as Fig. \ref{fg:SF_W0116} but for W0220+0137.}
    \label{fg:SF_W0220}
  \end{center}
\end{figure}

\begin{figure}
  \begin{center}
    \plotone{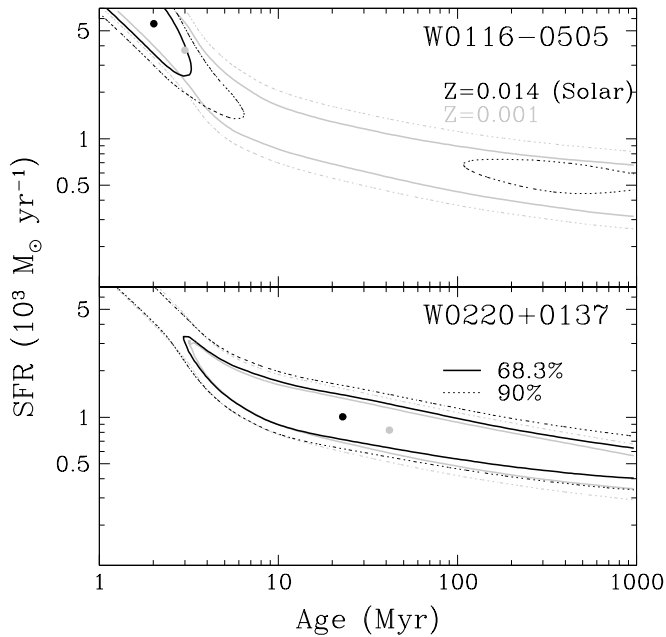}
    \caption{The contours show a $\chi^2$ map of the best-fit
      Starburst99 models as discussed in the text. The contours for
      W0116--0505 and W0220+0137 are shown in the top and bottom
      panels, respectively. Dark contours assume solar metallicity,
      while gray contours assume $Z=0.001$. The solid (dotted) contour
      shows the 68.3\% (90\%) confidence region, while the solid dots
      show the values of the best-fits models shown in Figures
      \ref{fg:SF_W0116} and \ref{fg:SF_W0220}.}
    \label{fg:SFR_age}
  \end{center}
\end{figure}

For W0116--0505 and W0220+0137 the situation is somewhat
different. The optical spectra, shown in Figures \ref{fg:W0116_spec}
and \ref{fg:W0220_spec}, show clear broad, high-ionization features
characteristic of quasars. The UV emission, while spatially extended,
is strongly concentrated in both objects (see discussion in
\S\ref{ssec:hst_data}), which is more consistent with the expectations
for the dual AGN or the leaked AGN light scenarios, instead of the
star-formation scenario. We model the broad-band SEDs of these objects
as starbursts as in A16, but with a small modification. As the
approach of A16 only models the continuum emission, we add to the
Starburst99 models the emission lines of the spectra presented in
\S\ref{sec:observations}. Each emission line was modeled using a
single Gaussian profiles after subtracting a local continuum estimated
with a linear interpolation from the continuum regions immediately
blue and red of the emission line in question. The only exception to
this are Ly$\alpha$ and N{\,\sc{v}}, for which the continuum was only
modeled using the continuum region immediately red of them as the blue
continuum is affected by Ly$\alpha$ forest absorption. A single
Gaussian profile is good model of the N\,{\sc{v}}, C\,{\sc{iv}},
He\,{\sc{ii}} and C\,{\sc iii}] emission as well as for the
  S\,{\sc{iv}}+O\,{\sc{iv}}] line complex. For Ly$\alpha$ we recover
    the majority of the flux in both objects, but the peak of the
    best-fit Gaussian falls somewhat short of their observed narrow
    peak. The best-fit models are shown in Figures \ref{fg:SF_W0116}
    and \ref{fg:SF_W0220} for W0116--0505 and W0220+0137,
    respectively. Each figure shows the best fit obtained assuming a
    solar metallicity and a metallicity of $Z=0.001$ discussed
    above. As we do not include additional nebular emission or
    Ly$\alpha$ forest absorption, we only use the bands that are
    redward of the Ly$\alpha$ emission line and exclude the F160W
    band, which can be strongly contaminated by an older stellar
    population and by nebular [O\,{\sc ii}] emission \citep[which
      could be broadened and have a high equivalent width, as found
      by][for a different group of Hot DOGs]{jun18}. Indeed, if we
    include the F160W band we find best-fit $\chi^2$ values a factor
    $\sim$3 larger. We set a minimum photometric uncertainty of
    0.05~mag as systematic differences between the measurements are
    unlikely to be below that level. In practice, this only affects
    the uncertainty used for the F555W band. The best-fit SFR, age and
    obscuration of the stellar population are shown in the figures as
    well; however, the values are quite degenerate as shown in Figure
    \ref{fg:SFR_age}, particularly as there are no constraints
    longwards of $\sim$4000\AA. The only other longer wavelength
    broadbands that we have are in regions of the SED dominated by
    either an older stellar population or by the highly obscured,
    luminous AGN as shown in Figure \ref{fg:seds}, and hence are not
    useful for constraining these fits. We note, however, that the
    $\chi^2$ values of the best fits are quite large when it is
    considered that we are fitting three different parameters. This,
    coupled with UV/optical spectral features (i.e., the presence of
    broad emission lines) and the morphology in the {\it{HST}} imaging
    suggest that the UV/optical emission in these objects is unlikely
    dominated by unobscured starbursts.

\subsection{Leaked AGN Light}

The third possibility to explain the nature of BHDs is that the blue
excess emission found in these objects corresponds to light coming
from the highly obscured primary AGN that is leaking into our line of
sight. As discussed by A16, this could happen either due to dust or
gas scattering of the AGN emission into our line of sight, or due to a
small gap in the dust that allows for a partial view towards the
accretion disk and the broad-line region. However, the latter is
unlikely, as discussed by A16, as the UV/optical SED is consistent
with the emission of an unobscured accretion disk. As radiation at
progressively shorter wavelengths is emitted in progressively inner
regions of the accretion disk, a gap that only allows $\sim$1\% of the
emitted light through but does not distort the accretion disk spectrum
would need to cover 99\% of the effective disk size at each
wavelength. While not impossible, the shape of such a gap would be
exceedingly contrived, making this unlikely. Furthermore, we would not
expect the UV/optical emission to be spatially extended, as found in
\S\ref{ssec:hst_data}.

This suggests that the most likely source of the blue excess emission
in the three BHDs studied is scattered light. In this scenario, 99\%
of the emission from the accretion disk and the broad-line region
would be absorbed by dust, while 1\% will be scattered into our line
of sight by either dust or gas, or both. Reflection nebulae are known
to make the reflected SED bluer than the emitted SED in the UV
\citep[$\lambda \lesssim 2500\rm \AA$, e.g. see][]{draine03a},
suggesting that the scattering medium in BHDs is more likely the gas
surrounding the AGN. However, as there is likely dust on scales larger
than the torus \citep{diaz16,tsai18}, we note that our data are not
sufficient to rule out an SED that has been made bluer by dust
reflection and redder by dust absorption. With respect to the X-rays,
we note that while they should also be scattered into our line of
sight along with the UV emission, the scattering cross section by
either dust or free electrons is significantly smaller at the high
energy ranges probed by the {\it{Chandra}} observations
\citep{draine03a,draine03b}. Hence, the non-detection of a clear
unabsorbed component in the X-ray spectra in \S\ref{sec:xray_model} is
consistent with this scattering scenario.

\section{Summary}\label{sec:summary}

We have investigated the source of the blue excess emission in three BHDs, two
of which were identified as such by A16, and a third one which has an SED
consistent with that of a Hot DOG although it does not meet the formal selection
criteria due to being slightly too bright in the W1 band. While all Hot DOGs are
characterized by mid-IR emission that is most naturally explained by a highly
obscured hyper-luminous AGN \citep{eisenhardt12,assef15,tsai15}, BHDs have a
UV/optical SED that is significantly bluer than expected based on template
fitting results. Using a similar approach to that of \citet{assef15}, we find
that the SEDs of BHDs are best modeled using two AGN components: a primary
hyper-luminous, highly obscured AGN that dominates the mid-IR emission, and a
secondary lower luminosity but unobscured AGN that dominates the UV/optical
emission. The bolometric luminosity of the secondary AGN SED is $\sim$1\% of
that of the primary component (although, as discussed at the end of
\S\ref{ssec:bhds}, there could be unaccounted systematic uncertainties in the
luminosities due to our minimalistic SED modeling approach). A16 identified
three possible scenarios to produce the excess blue emission, namely: (i) a
secondary, less luminous but unobscured AGN in the system, (ii) an extreme
starburst, or (iii) leaked UV/optical light from the primary, highly luminous,
highly obscured AGN that dominates the mid-IR.

For one of the sources (W0204--0506), A16 ruled out a secondary AGN as
the source of the blue excess emission, and instead concluded that the
excess was caused by either unobscured star formation with an ${\rm
  SFR}\gtrsim 1000~M_{\odot}~\rm yr^{-1}$, or by UV/optical light from
the central engine leaking into our line of sight due to scattering or
through a partially obscured sight-line, with the scattered AGN light
hypothesis deemed more likely.  In this paper, we have presented
{\it{HST}}/WFC3 imaging of W0204--0506 showing a morphology consistent
with an on-going merger and evidence of an on-going widespread
starburst. Considering, however, the very high SFR needed to explain
the UV emission by star-formation alone, we conclude it is more likely
that the UV emission of W0204--0505 arises from a combination of
scattered AGN light and star-formation.

We also studied in detail two other BHDs, W0116--0505 and
W0220+0137. We present observations obtained with
{\it{Chandra}}/ACIS-S and interpret them using the \citet{brightman11}
models. We find that the X-ray spectra are consistent with single
luminous, highly absorbed AGNs dominating the X-ray emission. We find
that the $L_{2-10~\rm keV}$ luminosities of these AGNs are consistent
with those expected for the primary AGNs based on their estimated
$L_{6\mu\rm m}$ according to the $L_{6\mu\rm m}-L_{2-10~\rm keV}$
relation of \citet{stern15}. We also find that the gas-to-dust ratios
of the AGNs in these systems are somewhat below the median value found
in AGNs by \citet{maiolino01} and lower than that found in
W0204--0506, suggestive of a lower metallicity or of a higher fraction
of absorbing gas within the dust-sublimation radius of the AGN. Based
on the UV through mid-IR SED models of these sources, we estimate the
expected X-ray luminosity of the putative secondary AGN components
assuming it is a second, independent AGN in the system. We found that
the X-ray observations are only marginally consistent with the
presence of a second AGN component in both W0116--0505 and W0220+0137,
suggesting the dual AGN scenario is unlikely.

We followed A16 and modeled the UV emission of W0116--0505 and
W0220+0137 assuming a pure starburst scenario, and found that while
the best-fit SFRs are generally high, comparable to those found by A16
for W0204--0506, they are not well constrained due to the large
degeneracies between SFR, age and metallicity. We found, however, that
the $\chi^{2}$ values of the best-fit starburst models are large
($\sim 12$ for W0116--0506 and $\sim 8$ for W0220+0137) despite the
small number of degrees of freedom (1 and 2 respectively), implying a
pure starburst is not a good description of the observed UV
SED. Additionally, the rest-frame UV spectra shows broad emission
lines characteristic of AGN activity, further suggesting that
star-formation does not dominate the observed UV emission.

Finally, we also studied the morphologies observed in the
{\it{HST}}/WFC3 F555W and F160W images of W0116--0505 and W0220+0137
and found them to be undisturbed with the UV emission being centrally
concentrated. An analysis based on the Gini, $M_{20}$ and $A$
coefficients showed that these systems are best characterized as
undisturbed early type galaxies, consistent with the leaked AGN
scenario. Considering all of this, we conclude that the source of the
UV emission in W0116--0505 and W0220+0137 is scattered light from the
hyperluminous, highly obscured AGN that powers the mid-IR SED. Given
the detail of our data and SED modeling, we cannot determine whether
the scattering material is primarily gas, dust, or a mixture of
both. As discussed in \S\ref{sec:xray_model}, W0220+0137 is classified
as an ERQ and W0116-0505 is classified as ERQ-like
\citep{hamann17}. Recently, \citet{alexandroff18} found high
polarization fractions in the UV spectra of 3 ERQs, implying the UV
emission has a strong contribution by scattered light from the central
engine for these objects. This is highly consistent with our
conclusions.

That all three BHDs we have investigated are due to scattered light
from the highly obscured, hyperluminous AGN highlights how powerful
the central engine is in Hot DOGs: with only 1\% of the emission of
the accretion disk scattered into our line of sight, it is still more
luminous than the entire stellar emission of the host galaxy in the
UV. This is in general agreement with recent results which show that
the SMBHs in Hot DOGs are accreting above the Eddington limit
\citep{wu18,tsai18} and are injecting large amounts of energy into the
ISM of their host galaxies \citep{diaz16}, and hence are experiencing
strong events of AGN feedback.

\acknowledgments

We thank J. Comerford and B. Weiner for carrying out observations presented in
this article. We also thank the anonymous referee for the comments and
suggestions to improve the manuscript.  RJA was supported by FONDECYT grants
number 1151408 and 1191124. DJW acknowledges financial support from STFC Ernest
Rutherford fellowships. HDJ was supported by Basic Science Research Program
through the National Research Foundation of Korea (NRF) funded by the Ministry
of Education (NRF-2017R1A6A3A04005158). FEB acknowledges support from
CONICYT-Chile (Basal AFB-170002) and the Ministry of Economy, Development and
Tourism's Millenium Science Initiative through grant IC120009, awarded to The
Millenium Institute of Astrophysics, MAS. T.D-S. acknowledges support from the
CASSACA and CONICYT fund CAS-CONICYT Call 2018. JW is supported by the NSFC
Grant 11690024 and SPRP CAS grant XDB23000000.

Based on observations made with the NASA/ESA {\it{Hubble Space Telescope}},
obtained at the Space Telescope Science Institute, which is operated by the
Association of Universities for Research in Astronomy, Inc., under NASA contract
NAS 5-26555. These observations are associated with program \#14358. Support for
program \#14358 was provided by NASA through a grant from the Space Telescope
Science Institute, which is operated by the Association of Universities for
Research in Astronomy, Inc., under NASA contract NAS 5-26555. The scientific
results reported in this article are based to a significant degree on data
obtained from the {\it{Chandra X-ray Observatory}} and observations made by the
{\it{Chandra X-ray Observatory}} and published previously in cited articles.
Support for this work was provided by the National Aeronautics and Space
Administration through Chandra Award Number 17700696 issued by the Chandra X-ray
Center, which is operated by the Smithsonian Astrophysical Observatory for and
on behalf of the National Aeronautics Space Administration under contract
NAS8-03060. This publication makes use of data products from the {\it{Wide-field
Infrared Survey Explorer}}, which is a joint project of the University of
California, Los Angeles, and the Jet Propulsion Laboratory/California Institute
of Technology, funded by the National Aeronautics and Space Administration. This
work is based in part on observations made with the {\it{Spitzer Space
Telescope}}, which is operated by the Jet Propulsion Laboratory, California
Institute of Technology under a contract with NASA. The Pan-STARRS1 Surveys
(PS1) have been made possible through contributions of the Institute for
Astronomy, the University of Hawaii, the Pan-STARRS Project Office, the
Max-Planck Society and its participating institutes, the Max Planck Institute
for Astronomy, Heidelberg and the Max Planck Institute for Extraterrestrial
Physics, Garching, The Johns Hopkins University, Durham University, the
University of Edinburgh, Queen's University Belfast, the Harvard-Smithsonian
Center for Astrophysics, the Las Cumbres Observatory Global Telescope Network
Incorporated, the National Central University of Taiwan, the Space Telescope
Science Institute, the National Aeronautics and Space Administration under Grant
No. NNX08AR22G issued through the Planetary Science Division of the NASA Science
Mission Directorate, the National Science Foundation under Grant No.
AST-1238877, the University of Maryland, and Eotvos Lorand University (ELTE) and
the Los Alamos National Laboratory. Funding for SDSS-III has been provided by
the Alfred P. Sloan Foundation, the Participating Institutions, the National
Science Foundation, and the U.S. Department of Energy Office of Science. The
SDSS-III web site is {\url{http:// www.sdss3.org/}}. SDSS-III is managed by the
Astrophysical Research Consortium for the Participating Institutions of the
SDSS-III Collaboration including the University of Arizona, the Brazilian
Participation Group, Brookhaven National Laboratory, Carnegie Mellon University,
University of Florida, the French Participation Group, the German Participation
Group, Harvard University, the Instituto de Astrofisica de Canarias, the
Michigan State/Notre Dame/JINA Participation Group, Johns Hopkins University,
Lawrence Berkeley National Laboratory, Max Planck Institute for Astrophysics,
Max Planck Institute for Extraterrestrial Physics, New Mexico State University,
New York University, Ohio State University, Pennsylvania State University,
University of Portsmouth, Princeton University, the Spanish Participation Group,
University of Tokyo, University of Utah, Vanderbilt University, University of
Virginia, University of Washington, and Yale University. Some of the
observations reported here were obtained at the MMT Observatory, a joint
facility of the Smithsonian Institution and the University of Arizona.


\begin{thebibliography}{}

\bibitem[{{Abraham} {et~al.}(2003){Abraham}, {van den Bergh}, \&
  {Nair}}]{abraham03}
{Abraham}, R.~G., {van den Bergh}, S., \& {Nair}, P. 2003, \apj, 588, 218

\bibitem[{{Alexander} \& {Hickox}(2012)}]{alexander12}
  {Alexander}, D.~M., \& {Hickox}, R.~C. 2012, \nar, 56, 93

\bibitem[Alexandroff et al.(2018)]{alexandroff18}
  Alexandroff, R.~M., Zakamska, N.~L., Barth, A.~J., et al.\ 2018,
  \mnras, 479, 4936

\bibitem[Alexandroff et al.(2013)]{alexandroff13}
  Alexandroff, R., Strauss, M.~A., Greene, J.~E., et al.\ 2013,
  \mnras, 435, 3306

\bibitem[{{Assef} {et~al.}(2010){Assef}, {Kochanek}, {Brodwin}, {Cool},
  {Forman}, {Gonzalez}, {Hickox}, {Jones}, {Le Floc'h}, {Moustakas}, {Murray},
  \& {Stern}}]{assef10}
{Assef}, R.~J., {Kochanek}, C.~S., {Brodwin}, M., {et~al.} 2010, \apj, 713, 970

\bibitem[{{Assef} {et~al.}(2011){Assef}, {Denney}, {Kochanek}, {Peterson},
  {Koz{\l}owski}, {Ageorges}, {Barrows}, {Buschkamp}, {Dietrich}, {Falco},
  {Feiz}, {Gemperlein}, {Germeroth}, {Grier}, {Hofmann}, {Juette}, {Khan},
  {Kilic}, {Knierim}, {Laun}, {Lederer}, {Lehmitz}, {Lenzen}, {Mall}, {Madsen},
  {Mandel}, {Martini}, {Mathur}, {Mogren}, {Mueller}, {Naranjo}, {Pasquali},
  {Polsterer}, {Pogge}, {Quirrenbach}, {Seifert}, {Stern}, {Shappee}, {Storz},
  {Van Saders}, {Weiser}, \& {Zhang}}]{assef11}
{Assef}, R.~J., {Denney}, K.~D., {Kochanek}, C.~S., {et~al.} 2011, \apj, 742,
  93

\bibitem[{{Assef} {et~al.}(2015){Assef}, {Eisenhardt}, {Stern}, {Tsai}, {Wu},
  {Wylezalek}, {Blain}, {Bridge}, {Donoso}, {Gonzales}, {Griffith}, \&
  {Jarrett}}]{assef15}
{Assef}, R.~J., {Eisenhardt}, P.~R.~M., {Stern}, D., {et~al.} 2015, \apj, 804,
  27

\bibitem[{{Assef} {et~al.}(2016){Assef}, {Walton}, {Brightman}, {Stern},
  {Alexander}, {Bauer}, {Blain}, {Diaz-Santos}, {Eisenhardt}, {Finkelstein},
  {Hickox}, {Tsai}, \& {Wu}}]{assef16}
{Assef}, R.~J., {Walton}, D.~J., {Brightman}, M., {et~al.} 2016, \apj, 819, 111

\bibitem[{{Banerji} {et~al.}(2015){Banerji}, {Alaghband-Zadeh}, {Hewett}, \&
  {McMahon}}]{banerji15}
{Banerji}, M., {Alaghband-Zadeh}, S., {Hewett}, P.~C., \& {McMahon}, R.~G.
  2015, \mnras, 447, 3368

\bibitem[{{Barger} {et~al.}(2014){Barger}, {Cowie}, {Chen}, {Owen}, {Wang},
  {Casey}, {Lee}, {Sanders}, \& {Williams}}]{barger14}
{Barger}, A.~J., {Cowie}, L.~L., {Chen}, C.-C., {et~al.} 2014, \apj, 784, 9

\bibitem[{{Bauer} {et~al.}(2010){Bauer}, {Yan}, {Sajina}, \&
  {Alexander}}]{bauer10}
{Bauer}, F.~E., {Yan}, L., {Sajina}, A., \& {Alexander}, D.~M. 2010, \apj, 710,
  212

\bibitem[{{Bennert} {et~al.}(2011){Bennert}, {Auger}, {Treu}, {Woo}, \&
  {Malkan}}]{bennert11}
{Bennert}, V.~N., {Auger}, M.~W., {Treu}, T., {Woo}, J.-H., \& {Malkan}, M.~A.
  2011, \apj, 726, 59

\bibitem[{{Bertin} \& {Arnouts}(1996)}]{bertin96}
{Bertin}, E., \& {Arnouts}, S. 1996, \aaps, 117, 393

\bibitem[{{Brightman} \& {Nandra}(2011)}]{brightman11}
{Brightman}, M., \& {Nandra}, K. 2011, \mnras, 413, 1206

\bibitem[{{Cash}(1979)}]{cash79}
{Cash}, W. 1979, \apj, 228, 939

\bibitem[{{Chen} {et~al.}(2017){Chen}, {Hickox}, {Goulding}, {Stern}, {Assef},
  {Kochanek}, {Brown}, {Harrison}, {Hainline}, {Alberts}, {Alexander},
  {Brodwin}, {Del Moro}, {Forman}, {Gorjian}, {Jones}, {Murray}, {Pope}, \&
  {Rovilos}}]{chen17}
{Chen}, C.-T.~J., {Hickox}, R.~C., {Goulding}, A.~D., {et~al.} 2017, \apj, 837,
  145

\bibitem[{{Conselice}(2014)}]{conselice14}
{Conselice}, C.~J. 2014, \araa, 52, 291

\bibitem[{{Cutri} {et~al.}(2012){Cutri}, {Wright}, {Conrow}, {Bauer},
  {Benford}, {Brandenburg}, {Dailey}, {Eisenhardt}, {Evans}, {Fajardo-Acosta},
  {Fowler}, {Gelino}, {Grillmair}, {Harbut}, {Hoffman}, {Jarrett},
  {Kirkpatrick}, {Leisawitz}, {Liu}, {Mainzer}, {Marsh}, {Masci}, {McCallon},
  {Padgett}, {Ressler}, {Royer}, {Skrutskie}, {Stanford}, {Wyatt}, {Tholen},
  {Tsai}, {Wachter}, {Wheelock}, {Yan}, {Alles}, {Beck}, {Grav}, {Masiero},
  {McCollum}, {McGehee}, {Papin}, \& {Wittman}}]{cutri12}
{Cutri}, R.~M., {Wright}, E.~L., {Conrow}, T., {et~al.} 2012, {Explanatory
  Supplement to the WISE All-Sky Data Release Products}, Tech. rep.

\bibitem[De Breuck et al.(2001)]{debreuck01}
  De Breuck, C., van Breugel, W., R{\"o}ttgering, H., et al.\ 2001,
  \aj, 121, 1241

\bibitem[{{D{\'{\i}}az-Santos} {et~al.}(2016){D{\'{\i}}az-Santos}, {Assef},
  {Blain}, {Tsai}, {Aravena}, {Eisenhardt}, {Wu}, {Stern}, \&
  {Bridge}}]{diaz16}
{D{\'{\i}}az-Santos}, T., {Assef}, R.~J., {Blain}, A.~W., {et~al.} 2016, \apjl,
  816, L6

\bibitem[{{D{\'{\i}}az-Santos} {et~al.}(2018){D{\'{\i}}az-Santos}, {Assef},
  {Blain}, {Aravena}, {Stern}, {Tsai}, {Eisenhardt}, {Wu}, {Jun}, {Dibert},
  {Inami}, {Lansbury}, \& {Leclercq}}]{diaz18}
---. 2018, Science, 362, 1034

\bibitem[{{Draine}(2003{\natexlab{a}})}]{draine03a}
{Draine}, B.~T. 2003{\natexlab{a}}, \apj, 598, 1017

\bibitem[{{Draine}(2003{\natexlab{b}})}]{draine03b}
---. 2003{\natexlab{b}}, \apj, 598, 1026

\bibitem[{{Eisenhardt} {et~al.}(2012){Eisenhardt}, {Wu}, {Tsai}, {Assef},
  {Benford}, {Blain}, {Bridge}, {Condon}, {Cushing}, {Cutri}, {Evans},
  {Gelino}, {Griffith}, {Grillmair}, {Jarrett}, {Lonsdale}, {Masci}, {Mason},
  {Petty}, {Sayers}, {Stanford}, {Stern}, {Wright}, \& {Yan}}]{eisenhardt12}
{Eisenhardt}, P.~R.~M., {Wu}, J., {Tsai}, C.-W., {et~al.} 2012, \apj, 755, 173

\bibitem[{{Fan} {et~al.}(2016{\natexlab{a}}){Fan}, {Han}, {Nikutta}, {Drouart},
  \& {Knudsen}}]{fan16a}
{Fan}, L., {Han}, Y., {Nikutta}, R., {Drouart}, G., \& {Knudsen}, K.~K.
  2016{\natexlab{a}}, \apj, 823, 107

\bibitem[{{Fan} {et~al.}(2016{\natexlab{b}}){Fan}, {Han}, {Fang}, {Gao},
  {Zhang}, {Jiang}, {Wu}, {Yang}, \& {Li}}]{fan16b}
{Fan}, L., {Han}, Y., {Fang}, G., {et~al.} 2016{\natexlab{b}}, \apjl, 822, L32

\bibitem[{{Farrah} {et~al.}(2017){Farrah}, {Petty}, {Connolly}, {Blain},
  {Efstathiou}, {Lacy}, {Stern}, {Lake}, {Jarrett}, {Bridge}, {Eisenhardt},
  {Benford}, {Jones}, {Tsai}, {Assef}, {Wu}, \& {Moustakas}}]{farrah17}
{Farrah}, D., {Petty}, S., {Connolly}, B., {et~al.} 2017, \apj, 844, 106

\bibitem[{{Finkelstein} {et~al.}(2007){Finkelstein}, {Rhoads}, {Malhotra},
  {Pirzkal}, \& {Wang}}]{finkelstein07}
{Finkelstein}, S.~L., {Rhoads}, J.~E., {Malhotra}, S., {Pirzkal}, N., \&
  {Wang}, J. 2007, \apj, 660, 1023

\bibitem[{{Fiore} {et~al.}(2009){Fiore}, {Puccetti}, {Brusa}, {Salvato},
  {Zamorani}, {Aldcroft}, {Aussel}, {Brunner}, {Capak}, {Cappelluti}, {Civano},
  {Comastri}, {Elvis}, {Feruglio}, {Finoguenov}, {Fruscione}, {Gilli},
  {Hasinger}, {Koekemoer}, {Kartaltepe}, {Ilbert}, {Impey}, {Le Floc'h},
  {Lilly}, {Mainieri}, {Martinez-Sansigre}, {McCracken}, {Menci}, {Merloni},
  {Miyaji}, {Sanders}, {Sargent}, {Schinnerer}, {Scoville}, {Silverman},
  {Smolcic}, {Steffen}, {Santini}, {Taniguchi}, {Thompson}, {Trump}, {Vignali},
  {Urry}, \& {Yan}}]{fiore09}
{Fiore}, F., {Puccetti}, S., {Brusa}, M., {et~al.} 2009, \apj, 693, 447

\bibitem[{{Gandhi} {et~al.}(2009){Gandhi}, {Horst}, {Smette}, {H{\"o}nig},
  {Comastri}, {Gilli}, {Vignali}, \& {Duschl}}]{ghandi09}
{Gandhi}, P., {Horst}, H., {Smette}, A., {et~al.} 2009, \aap, 502, 457

\bibitem[{{Goulding} {et~al.}(2018){Goulding}, {Zakamska}, {Alexandroff},
  {Assef}, {Banerji}, {Hamann}, {Wylezalek}, {Brandt}, {Greene}, {Lansbury},
  {P{\^a}ris}, {Richards}, {Stern}, \& {Strauss}}]{goulding18}
{Goulding}, A.~D., {Zakamska}, N.~L., {Alexandroff}, R.~M., {et~al.} 2018,
  \apj, 856, 4

\bibitem[{{Griffith} {et~al.}(2012){Griffith}, {Kirkpatrick}, {Eisenhardt},
  {Gelino}, {Cushing}, {Benford}, {Blain}, {Bridge}, {Cohen}, {Cutri},
  {Donoso}, {Jarrett}, {Lonsdale}, {Mace}, {Mainzer}, {Marsh}, {Padgett},
  {Petty}, {Ressler}, {Skrutskie}, {Stanford}, {Stern}, {Tsai}, {Wright}, {Wu},
  \& {Yan}}]{griffith12}
{Griffith}, R.~L., {Kirkpatrick}, J.~D., {Eisenhardt}, P.~R.~M., {et~al.} 2012,
  \aj, 144, 148

\bibitem[{{Hamann} {et~al.}(2017){Hamann}, {Zakamska}, {Ross}, {Paris},
  {Alexandroff}, {Villforth}, {Richards}, {Herbst}, {Brandt}, {Cook}, {Denney},
  {Greene}, {Schneider}, \& {Strauss}}]{hamann17}
{Hamann}, F., {Zakamska}, N.~L., {Ross}, N., {et~al.} 2017, \mnras, 464, 3431

\bibitem[Hainline et al.(2012)]{hainline12}
  Hainline, K.~N., Shapley, A.~E., Greene, J.~E., et al.\ 2012, \apj,
  760, 74

\bibitem[{{Hickox} \& {Alexander}(2018)}]{hickox18}
{Hickox}, R.~C., \& {Alexander}, D.~M. 2018, \araa, 56, 625

\bibitem[{{Hopkins} {et~al.}(2008){Hopkins}, {Hernquist}, {Cox}, \& {Kere{\v
  s}}}]{hopkins08}
{Hopkins}, P.~F., {Hernquist}, L., {Cox}, T.~J., \& {Kere{\v s}}, D. 2008,
  \apjs, 175, 356

\bibitem[{{Jun} {et~al.}(2020){Jun}, {Assef}, {Bauer}, {Blain},
    {Diaz-Santos}, {Eisenhardt}, {Stern}, {Tsai}, {Wright},
    {Wu}}]{jun18}
{Jun}, H.~D., {Assef}, R.~J., {Bauer}, F.~E., {et~al.} 2020, \apj,
accepted (arXiv:1911.09828)

\bibitem[{{Leitherer} {et~al.}(2014){Leitherer}, {Ekstr{\"o}m}, {Meynet},
  {Schaerer}, {Agienko}, \& {Levesque}}]{leitherer14}
{Leitherer}, C., {Ekstr{\"o}m}, S., {Meynet}, G., {et~al.} 2014, \apjs, 212, 14

\bibitem[{{Leitherer} {et~al.}(2010){Leitherer}, {Ortiz Ot{\'a}lvaro},
  {Bresolin}, {Kudritzki}, {Lo Faro}, {Pauldrach}, {Pettini}, \&
  {Rix}}]{leitherer10}
{Leitherer}, C., {Ortiz Ot{\'a}lvaro}, P.~A., {Bresolin}, F., {et~al.} 2010,
  \apjs, 189, 309

\bibitem[{{Leitherer} {et~al.}(1999){Leitherer}, {Schaerer}, {Goldader},
  {Delgado}, {Robert}, {Kune}, {de Mello}, {Devost}, \&
  {Heckman}}]{leitherer99}
{Leitherer}, C., {Schaerer}, D., {Goldader}, J.~D., {et~al.} 1999, \apjs, 123,
  3

\bibitem[{{Lotz} {et~al.}(2004){Lotz}, {Primack}, \& {Madau}}]{lotz04}
{Lotz}, J.~M., {Primack}, J., \& {Madau}, P. 2004, \aj, 128, 163

\bibitem[{{Lotz} {et~al.}(2008){Lotz}, {Davis}, {Faber}, {Guhathakurta},
  {Gwyn}, {Huang}, {Koo}, {Le Floc'h}, {Lin}, {Newman}, {Noeske}, {Papovich},
  {Willmer}, {Coil}, {Conselice}, {Cooper}, {Hopkins}, {Metevier}, {Primack},
  {Rieke}, \& {Weiner}}]{lotz08}
{Lotz}, J.~M., {Davis}, M., {Faber}, S.~M., {et~al.} 2008, \apj, 672, 177

\bibitem[{{Lupton} {et~al.}(2004){Lupton}, {Blanton}, {Fekete}, {Hogg},
  {O'Mullane}, {Szalay}, \& {Wherry}}]{lupton04}
{Lupton}, R., {Blanton}, M.~R., {Fekete}, G., {et~al.} 2004, \pasp, 116, 133

\bibitem[{{Magorrian} {et~al.}(1998){Magorrian}, {Tremaine}, {Richstone},
  {Bender}, {Bower}, {Dressler}, {Faber}, {Gebhardt}, {Green}, {Grillmair},
  {Kormendy}, \& {Lauer}}]{magorrian98}
{Magorrian}, J., {Tremaine}, S., {Richstone}, D., {et~al.} 1998, \aj, 115, 2285

\bibitem[{{Maiolino} {et~al.}(2001){Maiolino}, {Marconi}, {Salvati},
  {Risaliti}, {Severgnini}, {Oliva}, {La Franca}, \& {Vanzi}}]{maiolino01}
{Maiolino}, R., {Marconi}, A., {Salvati}, M., {et~al.} 2001, \aap, 365, 28

\bibitem[{{Mancone} \& {Gonzalez}(2012)}]{mancone12}
{Mancone}, C.~L., \& {Gonzalez}, A.~H. 2012, \pasp, 124, 606

\bibitem[{{Mateos} {et~al.}(2015){Mateos}, {Carrera}, {Alonso-Herrero},
  {Rovilos}, {Hern{\'a}n-Caballero}, {Barcons}, {Blain}, {Caccianiga}, {Della
  Ceca}, \& {Severgnini}}]{mateos15}
{Mateos}, S., {Carrera}, F.~J., {Alonso-Herrero}, A., {et~al.} 2015, \mnras,
  449, 1422

\bibitem[{{Petrosian}(1976)}]{petrosian76}
{Petrosian}, V. 1976, \apjl, 209, L1

\bibitem[{{Piconcelli} {et~al.}(2015){Piconcelli}, {Vignali}, {Bianchi},
  {Zappacosta}, {Fritz}, {Lanzuisi}, {Miniutti}, {Bongiorno}, {Feruglio},
  {Fiore}, \& {Maiolino}}]{piconcelli15}
{Piconcelli}, E., {Vignali}, C., {Bianchi}, S., {et~al.} 2015, \aap, 574, L9

\bibitem[{{Rhodes} {et~al.}(2000){Rhodes}, {Refregier}, \& {Groth}}]{rhodes00}
{Rhodes}, J., {Refregier}, A., \& {Groth}, E.~J. 2000, \apj, 536, 79

\bibitem[{{Ricci} {et~al.}(2017){Ricci}, {Assef}, {Stern}, {Nikutta},
  {Alexander}, {Asmus}, {Ballantyne}, {Bauer}, {Blain}, {Boggs}, {Boorman},
  {Brandt}, {Brightman}, {Chang}, {Chen}, {Christensen}, {Comastri}, {Craig},
  {D{\'{\i}}az-Santos}, {Eisenhardt}, {Farrah}, {Gandhi}, {Hailey}, {Harrison},
  {Jun}, {Koss}, {LaMassa}, {Lansbury}, {Markwardt}, {Stalevski}, {Stanley},
  {Treister}, {Tsai}, {Walton}, {Wu}, {Zappacosta}, \& {Zhang}}]{ricci17}
{Ricci}, C., {Assef}, R.~J., {Stern}, D., {et~al.} 2017, \apj, 835, 105

\bibitem[{{Stern}(2015)}]{stern15}
{Stern}, D. 2015, \apj, 807, 129

\bibitem[{{Stern} {et~al.}(2014){Stern}, {Lansbury}, {Assef}, {Brandt},
  {Alexander}, {Ballantyne}, {Balokovi{\'c}}, {Bauer}, {Benford}, {Blain},
  {Boggs}, {Bridge}, {Brightman}, {Christensen}, {Comastri}, {Craig}, {Del
  Moro}, {Eisenhardt}, {Gandhi}, {Griffith}, {Hailey}, {Harrison}, {Hickox},
  {Jarrett}, {Koss}, {Lake}, {LaMassa}, {Luo}, {Tsai}, {Urry}, {Walton},
  {Wright}, {Wu}, {Yan}, \& {Zhang}}]{stern14}
{Stern}, D., {Lansbury}, G.~B., {Assef}, R.~J., {et~al.} 2014, \apj, 794, 102

\bibitem[Stern et al.(2002)]{stern02}
  Stern, D., Moran, E.~C., Coil, A.~L., et al.\ 2002, \apj, 568, 71

\bibitem[Stern et al.(1999)]{stern99}
  Stern, D., Dey, A., Spinrad, H., et al.\ 1999, \aj, 117, 1122

\bibitem[{{Tsai} {et~al.}(2015){Tsai}, {Eisenhardt}, {Wu}, {Stern}, {Assef},
  {Blain}, {Bridge}, {Benford}, {Cutri}, {Griffith}, {Jarrett}, {Lonsdale},
  {Masci}, {Moustakas}, {Petty}, {Sayers}, {Stanford}, {Wright}, {Yan},
  {Leisawitz}, {Liu}, {Mainzer}, {McLean}, {Padgett}, {Skrutskie}, {Gelino},
  {Beichman}, \& {Juneau}}]{tsai15}
{Tsai}, C.-W., {Eisenhardt}, P.~R.~M., {Wu}, J., {et~al.} 2015, \apj, 805, 90

\bibitem[{{Tsai} {et~al.}(2018){Tsai}, {Eisenhardt}, {Jun}, {Wu}, {Assef},
  {Blain}, {D{\'{\i}}az-Santos}, {Jones}, {Stern}, {Wright}, \& {Yeh}}]{tsai18}
{Tsai}, C.-W., {Eisenhardt}, P.~R.~M., {Jun}, H.~D., {et~al.} 2018, \apj, 868,
  15

\bibitem[{{Tsai} {et~al.}(in prep.)}]{tsai20}
{Tsai}, C.-W., et~al.\ in prep.

\bibitem[{{van Dokkum}(2001)}]{vandokkum01}
{van Dokkum}, P.~G. 2001, \pasp, 113, 1420

\bibitem[{{V{\'a}zquez} \& {Leitherer}(2005)}]{vazquez05}
{V{\'a}zquez}, G.~A., \& {Leitherer}, C. 2005, \apj, 621, 695

\bibitem[{{Vito} {et~al.}(2018){Vito}, {Brandt}, {Stern}, {Assef}, {Chen},
  {Brightman}, {Comastri}, {Eisenhardt}, {Garmire}, {Hickox}, {Lansbury},
  {Tsai}, {Walton}, \& {Wu}}]{vito18}
{Vito}, F., {Brandt}, W.~N., {Stern}, D., {et~al.} 2018, \mnras, 474, 4528

\bibitem[{{Wright} {et~al.}(2010){Wright}, {Eisenhardt}, {Mainzer}, {Ressler},
  {Cutri}, {Jarrett}, {Kirkpatrick}, {Padgett}, {McMillan}, {Skrutskie},
  {Stanford}, {Cohen}, {Walker}, {Mather}, {Leisawitz}, {Gautier}, {McLean},
  {Benford}, {Lonsdale}, {Blain}, {Mendez}, {Irace}, {Duval}, {Liu}, {Royer},
  {Heinrichsen}, {Howard}, {Shannon}, {Kendall}, {Walsh}, {Larsen}, {Cardon},
  {Schick}, {Schwalm}, {Abid}, {Fabinsky}, {Naes}, \& {Tsai}}]{wright10}
{Wright}, E.~L., {Eisenhardt}, P.~R.~M., {Mainzer}, A.~K., {et~al.} 2010, \aj,
  140, 1868

\bibitem[{{Wu} {et~al.}(2012){Wu}, {Tsai}, {Sayers}, {Benford}, {Bridge},
  {Blain}, {Eisenhardt}, {Stern}, {Petty}, {Assef}, {Bussmann}, {Comerford},
  {Cutri}, {Evans}, {Griffith}, {Jarrett}, {Lake}, {Lonsdale}, {Rho},
  {Stanford}, {Weiner}, {Wright}, \& {Yan}}]{wu12}
{Wu}, J., {Tsai}, C.-W., {Sayers}, J., {et~al.} 2012, \apj, 756, 96

\bibitem[{{Wu} {et~al.}(2014){Wu}, {Bussmann}, {Tsai}, {Petric}, {Blain},
  {Eisenhardt}, {Bridge}, {Benford}, {Stern}, {Assef}, {Gelino}, {Moustakas},
  \& {Wright}}]{wu14}
{Wu}, J., {Bussmann}, R.~S., {Tsai}, C.-W., {et~al.} 2014, \apj, 793, 8

\bibitem[{{Wu} {et~al.}(2018){Wu}, {Jun}, {Assef}, {Tsai}, {Wright},
  {Eisenhardt}, {Blain}, {Stern}, {D{\'{\i}}az-Santos}, {Denney}, {Hayden},
  {Perlmutter}, {Aldering}, {Boone}, \& {Fagrelius}}]{wu18}
{Wu}, J., {Jun}, H.~D., {Assef}, R.~J., {et~al.} 2018, \apj, 852, 96

\bibitem[{{Yan} {et~al.}(2019){Yan}, {Hickox}, {Hainline}, {Stern}, {Lansbury},
  {Alexander}, {Hviding}, {Assef}, {Ballantyne}, {Dipompeo}, {Lanz}, {Carroll},
  {Koss}, {Lamperti}, {Civano}, {Del Moro}, {Gandhi}, \& {Myers}}]{yan19}
{Yan}, W., {Hickox}, R.~C., {Hainline}, K.~N., {et~al.} 2019, \apj, 870, 33

\bibitem[{{Zakamska} {et~al.}(2016){Zakamska}, {Hamann}, {P{\^a}ris}, {Brandt},
  {Greene}, {Strauss}, {Villforth}, {Wylezalek}, {Alexandroff}, \&
  {Ross}}]{zakamska16}
{Zakamska}, N.~L., {Hamann}, F., {P{\^a}ris}, I., {et~al.} 2016, \mnras, 459,
  3144

\end{thebibliography}
\end{document}